\documentclass[aps,10pt,prd,floatfix,twocolumn,nofootinbib,superscriptaddress,groupedaddress,showkeys]{revtex4-2}

\usepackage{amsmath,amssymb,bm,color,graphicx,mathrsfs}
\usepackage{float}
\usepackage{hyperref}
\usepackage{Macro}
\hypersetup{colorlinks=true,linkcolor=blue,citecolor=blue,urlcolor=blue}

\usepackage{cancel}
\def\hT{\hat{\Theta}}
\def\oT{\bar{\Theta}}
\def\T{\Theta}
\def\bu{\underline{\bl}}

\def\CTT{C^{\T\T}}
\def\oCTT{C^{\oT\oT}}

\def\estk{\hat{\kappa}}
\def\Planck{{\it Planck}}
\newcommand{\tl}{\tilde{\l}}
\newcommand{\tm}{\tilde{m}}

\begin{document}

\title{Faster CMB lensing with control variates}

\author{Toshiya Namikawa}
\affiliation{Kavli IPMU (WPI), UTIAS, The University of Tokyo, Kashiwa, 277-8583, Japan}

\author{Blake D. Sherwin}
\affiliation{Department of Applied Mathematics and Theoretical Physics, University of Cambridge, Wilberforce Road, Cambridge CB3 0WA, United Kingdom}
\affiliation{Kavli Institute for Cosmology, University of Cambridge, Madingley Road, Cambridge CB3 0HA, United Kingdom}

\date{\today}

\begin{abstract}
We present a new method for fast computation of the realization-dependent bias, a major computational bottleneck in measurements of the cosmic microwave background (CMB) lensing power spectrum. The method accelerates the bias calculation by differencing two correlated estimates: one based on fully realistic masked simulations and the other on isotropic simulations, for which the bias is analytically tractable. We show that our algorithm reduces the total computational cost of a lensing power spectrum measurement by approximately a factor of five for Atacama Cosmology Telescope- or Simons Observatory-like noise levels, or by a factor of three if current anisotropic filtering methods are left unchanged. Owing to its simplicity, the method can be readily implemented in existing CMB lensing analysis pipelines.
\end{abstract}



\maketitle

\section{Introduction} \label{sec:intro}

Gravitational lensing of the cosmic microwave background (CMB) encodes the projected matter distribution between the surface of last scattering and the observer, and is therefore a powerful probe of the growth of structure and the geometry of the Universe \cite{Blanchard:1987AA,Lewis:2006:review}. Its angular power spectrum has been measured with high precision by the Atacama Cosmology Telescope (ACT) \cite{Das:2011,Sherwin:2011:DE,ACT16:phi,Qu:ACT:2023,MacCrann:ACT:2023,Abril-Cabezas:2025}, BICEP/{\it Keck Array} \cite{BKVIII,BICEPKeck:2022:los-dist}, \Planck\ \cite{PR1:phi,PR2:phi,Carron:2022:NPIPE-lensing}, POLARBEAR \cite{PB:phi:2013,PB:phi:2019}, and the South Pole Telescope (SPT) \cite{SPT:phi:2012,Story:2015,SPT:phi:2019,Millea:2020,Pan:SPT:2023}, as well as through joint analyses combining data from multiple CMB experiments \cite{Omori:2017:SPT+Planck,Qu:2025:ACT+Planck+SPT}. These measurements have been used to constrain dark energy and modified gravity, the sum of neutrino masses, and the growth of structure in the late-time Universe; see, e.g., \cite{Sherwin:2011:DE,PR2:DE,PR3:main,Madhavacheril:ACT:2023,Madhavacheril:2024:review,Chudaykin:2025gdn}. Upcoming and future ground- and space-based CMB experiments, including the Simons Observatory \cite{SimonsObservatory,SimonsObservatory:LAT}, the Ali CMB Polarization Telescope (AliCPT) \cite{AliCPT:phi}, and LiteBIRD \cite{LiteBIRD,LiteBIRD:2023phi,LiteBIRD:2025:phi}, aim to provide increasingly sensitive measurements of CMB polarization, further improving CMB lensing reconstruction. Recent mild tension between CMB and baryon acoustic oscillation observations may also be interpreted as evidence for a CMB lensing amplitude larger than expected in the standard cosmological model \cite{Ge:SPT-3G:2024,Jhaveri:2025:tau}. Accurate determination of the lensing amplitude from CMB lensing observations is therefore crucial for robust cosmological inference.

Techniques for measuring the CMB lensing power spectrum have improved substantially over the past two decades, but this progress has been accompanied by increasing computational cost. Most CMB lensing measurements have relied on the quadratic estimator \cite{Hu:2001:cmbrecon,OkamotoHu:quad}, in which the reconstructed lensing map is quadratic in the observed CMB anisotropies. The CMB lensing power spectrum is then obtained from the power spectrum of the reconstructed fields after subtracting non-lensing bias contributions. Among these contributions, the dominant term is the disconnected part of the four-point correlation function of the CMB anisotropies, namely the Gaussian part. This disconnected, Gaussian bias can be interpreted as the power spectrum of the lensing reconstruction noise; it remains non-zero even in the limit of vanishing instrumental noise because of CMB cosmic variance.

Evaluating this disconnected bias is one of the most computationally expensive steps in CMB lensing analyses. An accurate and optimal estimate can be obtained using the realization-dependent approach, hereafter RDN0 \cite{Namikawa:2012:bhe}. However, for each data realization, RDN0 requires evaluating the disconnected bias using a large number of Monte Carlo (MC) simulations \cite{PR1:phi,PR2:phi,PR3:phi,Qu:ACT:2023}. For a single measurement, achieving sufficient convergence of this MC estimate typically requires several hundred simulations in recent CMB analyses. As a result, the RDN0 calculation is often the dominant computational bottleneck in CMB lensing analyses and can take hundreds of times longer than a typical lensing map reconstruction. Its high computational cost also makes it impractical to repeat the full RDN0 calculation for every mock realization when estimating the lensing spectrum covariance, and approximate treatments are often adopted.

In this paper, we propose a method for accelerating the computation of RDN0, and hence CMB lensing power spectrum measurements. Our approach relies on the method of control variates, which was first developed in statistics \cite{KahnMarshall:1953,LavenbergWelch:1981} and has recently been introduced to cosmological applications \cite{Chartier:2020:carpool,Kokron:2022,Ding2022DESICARPool,Chartier:2022:carpool,DeRose2023ZeldovichControlVariatesRSD,Hadzhiyska2023AnalyticControlVariatesDESI,Doytcheva2024HydroControlVariates,Hadzhiyska2025LyalphaBAOControlVariates,Kokron2025BispectrumControlVariates}. The algorithm employs two sets of correlated MC simulations: a fully realistic simulation and an auxiliary simulation with an isotropic sky signal. Although RDN0 depends on the anisotropic data, the expectation value of the RDN0 contribution from the isotropic auxiliary simulation is known analytically, allowing it to be used as a control variate. By differencing the RDN0 estimates obtained from the two simulations and adding back the analytic expectation value of the isotropic contribution, we construct an estimator that significantly reduces the computational cost of the standard RDN0 calculation. This method is intended for lensing analyses using data splits, which avoid noise correlations in power spectrum estimation \cite{Hanson:2009:noise,Madhavacheril:2020:split}.

The paper is organized as follows. In Sec.~\ref{sec:cmblensing}, we review the standard CMB lensing reconstruction method based on the quadratic estimator. In Sec.~\ref{sec:control-variate}, we introduce the proposed control-variate method for fast RDN0 computation and demonstrate the algorithm using simulations. Finally, in Sec.~\ref{sec:summary}, we summarize our findings and discuss the implications of the method.

\section{CMB lensing power spectrum estimation} \label{sec:cmblensing}

In this section, we review how the CMB lensing power spectrum is estimated from CMB observations. We focus on the quadratic estimator constructed from CMB temperature anisotropies.

\subsection{Estimating the CMB lensing power spectrum}

The lensed CMB temperature anisotropies, $\tilde{\T}$, are described as a remapping of the unlensed CMB anisotropies at the last-scattering surface, $\T$, by the displacement vector $\bm{d}(\hatn)$:
\al{
    \tilde{\T}(\hatn)=\T(\hatn+\bm{d}(\hatn))
    \,,
}
where $\hatn$ is a line-of-sight unit vector. See, e.g., Refs.~\cite{Lewis:2006:review,Hanson:2009:review,Bianchini:2025:review} for reviews. The displacement vector is related to the lensing convergence by $\kappa=-\bn\cdot\bm{d}/2$, where $\bn$ denotes the covariant derivative on the sphere. Expanding the lensed anisotropies to first order in $\bm{d}(\hatn)$ shows that, for a fixed lensing deflection field, lensing induces off-diagonal correlations between CMB temperature modes. This motivates the following temperature quadratic estimator for the harmonic coefficients of the lensing convergence \cite{OkamotoHu:quad}:
\al{
    \estk^*_{LM}[\oT,\oT'] = \frac{1}{2}A_L \sum_{\l\l'mm'} \Wjm{\l}{\l'}{L}{m}{m'}{M} f_{\l L\l'}\, \oT_{\l m}\oT'_{\l'm'}
    \,. \label{Eq:estg}
}
Here, the superscript $^*$ denotes complex conjugation, $f_{\l L\l'}$ is the weight function for the temperature quadratic estimator defined in Refs.~\cite{OkamotoHu:quad,Lewis:2011:bispec}, and $\oT_{\l m}$ and $\oT'_{\l' m'}$ are inverse-variance filtered temperature multipoles. In Eq.~\eqref{Eq:estg}, we keep the two filtered temperature fields distinct to facilitate the discussion below. For the lensing reconstruction, we set $\oT_{\l m}=\oT'_{\l m}$. The inverse-variance filtered temperature multipoles are defined by
\al{
    \oT_{\l m} = \{\bR{C}^{-1}\bm{\hT}\}_{\l m}
    \,,
}
where $\bm{\hT}$ is the vector of observed temperature multipoles, $\hT_{\l m}$, and $\bR{C}$ is their covariance matrix. In the following, we assume diagonal filtering, for which
\al{
    \oT_{\l m}=\frac{\hT_{\l m}}{\hCTT_{\l}}
    \,,
}
where $\hCTT_\l$ is the angular power spectrum of the observed temperature anisotropies.
\footnote{The method introduced in this paper, however, does not require diagonal filtering for real data or for the full simulations used to evaluate the conventional RDN0.}
The normalization factor $A_L$ is given by
\al{
    A^{-1}_L =
    \frac{1}{2L+1}
    \sum_{\l\l'}
    \frac{|f_{\l L\l'}|^2}{2\hCTT_{\l}\hCTT_{\l'}}
    \,.
}
The lensing convergence power spectrum is commonly measured from the power spectrum of the lensing estimator defined in Eq.~\eqref{Eq:estg}. Since the lensing estimator is quadratic in the observed temperature multipoles, its power spectrum corresponds to a four-point correlation function of the temperature anisotropies. It can be written as
\al{
    C_L^{\estk\estk} = \frac{1}{2L+1}\sum_M\ave{\estk_{LM}\estk^*_{LM}} = C^{\kappa\kappa}_L + N_L
    \,, \label{Eq:CL-estk}
}
where $\ave{\cdots}$ denotes an ensemble average, $C_L^{\kappa\kappa}$ is the lensing convergence power spectrum, and $N_L$ denotes bias and noise contributions to the estimator power spectrum. The dominant contribution to $N_L$ is the disconnected part of the four-point correlation function of the observed temperature anisotropies, commonly referred to as the ``N0 bias'', ``Gaussian bias'', or ``disconnected bias''. In the following, we focus on this disconnected contribution, which is the relevant term for the RDN0 calculation considered in this work. The lensing convergence power spectrum is then estimated by subtracting the bias terms from the power spectrum of the lensing estimator.

Since this bias arises from the disconnected part of the four-point temperature correlation, a straightforward way to estimate it is to use pairs of realizations drawn from two independent simulation sets, $S$ and $S'$. This gives
\al{
    N^{\rm MC}_L &= \ave{N^{{\rm MC},S_iS'_i}_L}_{i}
    \,, \label{Eq:MCN0}
}
where
\al{
    N^{{\rm MC},S_iS'_i}_L
    &=
    \frac{1}{2L+1}\sum_M
    \bigg[
    \left|\estk_{LM}[\oT^{S_i},\oT^{S'_i}]\right|^2
    \notag \\
    &\qquad +
    \estk_{LM}[\oT^{S_i},\oT^{S'_i}]
    \estk^*_{LM}[\oT^{S'_i},\oT^{S_i}]
    \bigg]
    \,. \label{Eq:RDN0:MC}
}
Here, $i$ labels the realization pairs and $\ave{\cdots}_i$ denotes an average over these pairs. Hereafter, we refer to $N^{\rm MC}_L$ as MCN0.

A more accurate, realization-dependent estimate of the disconnected bias is RDN0, which combines the data with simulations. This estimator is given by
\al{
    N^{\rm RD}_L
    = \ave{N^{{\rm DS},S_i}_L}_i - N^{\rm MC}_L
    \,, 
    \label{Eq:RDN0:full}
}
where the data--simulation cross term, $N^{{\rm DS},S_i}_L$, is defined as
\al{
    N^{{\rm DS},S_i}_L
    &\equiv
    \frac{1}{2L+1}\sum_M
    \bigg[\estk_{LM}[\oT^{S_i},\oT] \estk^*_{LM}[\oT^{S_i},\oT]
    \notag \\
    &\qquad
    + \estk_{LM}[\oT^{S_i},\oT] \estk^*_{LM}[\oT,\oT^{S_i}]
    \notag \\
    &\qquad
    + \estk_{LM}[\oT,\oT^{S_i}] \estk^*_{LM}[\oT^{S_i},\oT]
    \notag \\
    &\qquad
    + \estk_{LM}[\oT,\oT^{S_i}] \estk^*_{LM}[\oT,\oT^{S_i}]
    \bigg]
    \,.
    \label{Eq:RDN0:DS}
}
If the simulations have the same covariance of the temperature anisotropies as the data, the expectation value of the data--simulation cross term is $2N_L^{\rm MC}$, and hence $N^{\rm RD}_L=N^{\rm MC}_L$ in expectation. In practice, many MC realizations, typically several hundred, are required for the RDN0 estimate to converge. Moreover, the data--simulation cross term must be recomputed for each data realization or data split considered in the analysis. This repeated evaluation constitutes a major computational bottleneck in CMB lensing measurements. Reducing the cost of the RDN0 calculation is therefore crucial for practical CMB lensing analyses.

\subsection{Diagonal covariance approximation}

One might consider approximating RDN0 in order to reduce its computational cost. For example, consider an idealized situation in which the covariance of the filtered temperature anisotropies is diagonal:
\al{
    \ave{\oT_{\l m}\oT^*_{\l'm'}} = \oCTT_\l\delta_{\l\l'}\delta_{mm'}
    \,. 
}
Substituting Eq.~\eqref{Eq:estg} into Eq.~\eqref{Eq:MCN0} and using the above relation, MCN0 can be approximated as
\al{
    &N^{\rm MC}_L 
    \simeq \frac{1}{2L+1}\sum_M\frac{A_L^2}{4} \sum_{\l\l'mm'}\Wjm{\l}{\l'}{L}{m}{m'}{M}f_{\l L\l'}
    \notag \\
    &\qquad \times \sum_{\tl\tl'\tm\tm'}\Wjm{\tl}{\tl'}{L}{\tm}{\tm'}{M}f_{\tl L\tl'}
    \oCTT_\l\oCTT_{\l'}
    \notag \\
    &\qquad \times (\delta_{\l\tl}\delta_{m\tm}\delta_{\l'\tl'}\delta_{m'\tm'}+\delta_{\l\tl'}\delta_{m\tm'}\delta_{\l'\tl}\delta_{m'\tm})
    \notag \\
    &= \frac{A_L^2}{2L+1}\sum_{\l\l'}\frac{1}{2}f^2_{\l L\l'}\oCTT_\l\oCTT_{\l'}
    \,, 
}
where we have used $f_{\l L\l'}=f_{\l'L\l}$, $\l+L+\l'=$ even, and
\al{
    \sum_{mm'M}\Wjm{\l}{\l'}{L}{m}{m'}{M}\Wjm{\l}{\l'}{L}{m}{m'}{M} = 1
    \,. \label{Eq:Wigner-1}
}
The data--simulation contribution, $\ave{N^{{\rm DS},S_i}_L}_i$, can be approximated in a similar way. For example, the first term in Eq.~\eqref{Eq:RDN0:DS} becomes
\al{
    &\ave{N^{{\rm DS},S_i}_L}_i 
    \ni \frac{1}{2L+1}\sum_M\frac{A_L^2}{4} \sum_{\l\l'mm'}\Wjm{\l}{\l'}{L}{m}{m'}{M}f_{\l L\l'}
    \notag \\
    &\quad\times\sum_{\tl\tl'\tm\tm'}\Wjm{\tl}{\tl'}{L}{\tm}{\tm'}{M}f_{\tl L\tl'}\ave{\oT^{S_i}_{\l m}\oT_{\l'm'}(\oT^{S_i}_{\tl\tm})^*\oT^*_{\tl'\tm'}}_i
    \notag \\
    &\quad\simeq \frac{A_L^2}{2L+1} \sum_{\l\l'}\frac{f_{\l L\l'}^2}{4}\oCTT_\l\frac{1}{2\l'+1}\sum_{m'}\oT_{\l'm'}\oT^*_{\l'm'}
    \notag \\
    &\quad= \frac{A_L^2}{2L+1} \sum_{\l\l'}\frac{f^2_{\l L\l'}}{4}\oCTT_\l \widehat{C}^{\oT\oT}_{\l'}
    \,, 
}    
where we have used
\al{
    \sum_{mM}\Wjm{\l}{\l'}{L}{m}{m'}{M}\Wjm{\l}{\tl'}{L}{m}{\tm'}{M} = \frac{\delta_{\l'\tl'}\delta_{m'\tm'}}{2\l+1}
    \,, \label{Eq:Wigner-2}
}
and defined
\al{
    \widehat{C}^{\oT\oT}_{\l} \equiv \frac{1}{2\l+1}\sum_{m}\oT_{\l m}\oT^*_{\l m}
    \,. 
}
Applying the same approximation to the other terms in Eq.~\eqref{Eq:RDN0:DS}, the RDN0 expression in Eq.~\eqref{Eq:RDN0:full} becomes, under the diagonal approximation \cite{Hanson:2010:N2},
\al{
    &N^{\rm RD}_L \simeq \frac{A_L^2}{2L+1} \sum_{\l\l'}\frac{f^2_{\l L\l'}}{2}
    \notag \\
    &\quad\times \left(\widehat{C}^{\oT\oT}_{\l}\oCTT_{\l'}+\oCTT_{\l}\widehat{C}^{\oT\oT}_{\l'}-\oCTT_\l\oCTT_{\l'}\right)
    \,. \label{Eq:diag-RDN0}
}
The computational cost of this ``diagonal'' RDN0 is much lower than that of the full RDN0 calculation, because Eq.~\eqref{Eq:diag-RDN0} does not require repeated lensing reconstructions. However, diagonal RDN0 is useful only when the data are close to the idealized diagonal-covariance limit, i.e., when the effects of masks, inhomogeneous noise, and other sources of statistical anisotropy can be neglected. It therefore cannot generally be used in practical analyses as a replacement for the full RDN0 computation. This motivates a method that reduces the cost of the full RDN0 calculation without relying on the diagonal-covariance approximation.

\section{A control-variate method for RDN0} \label{sec:control-variate}

\begin{figure*}
\bc
\includegraphics[width=8.5cm,clip]{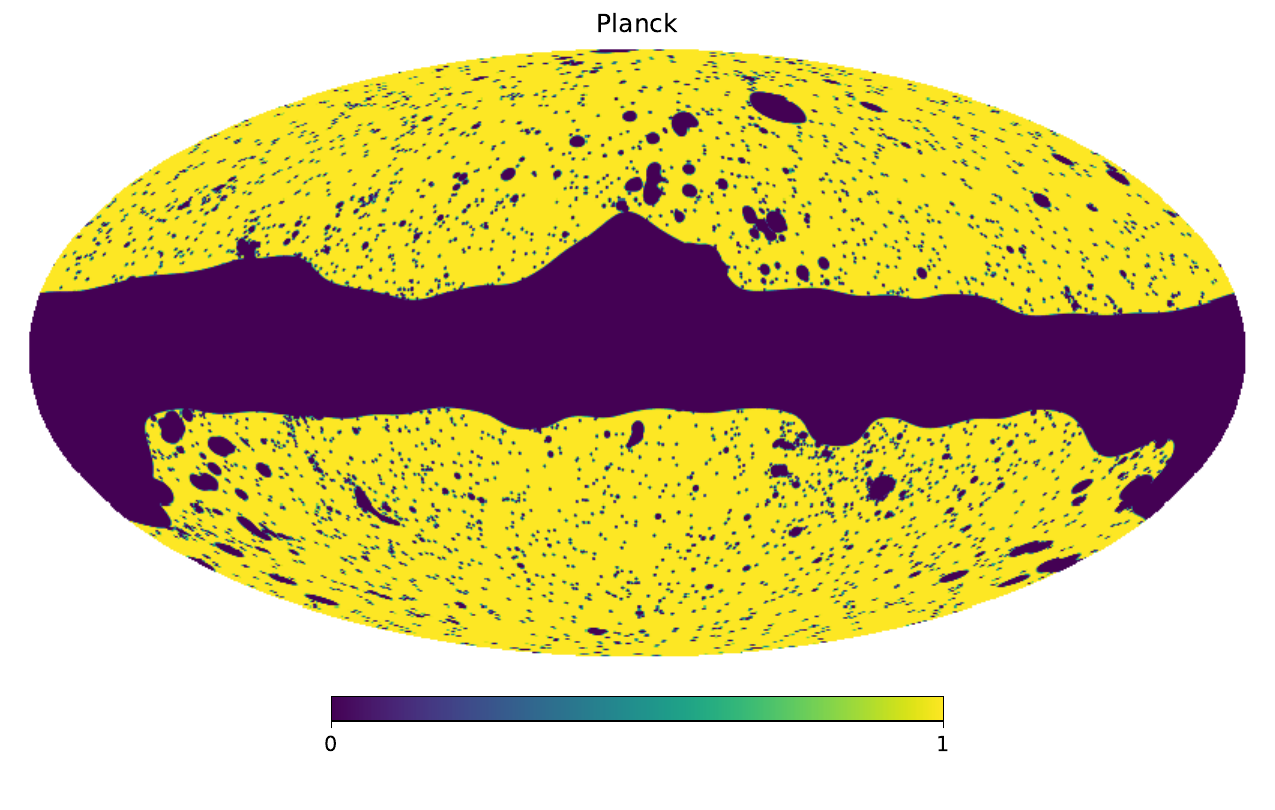}
\includegraphics[width=8.5cm,clip]{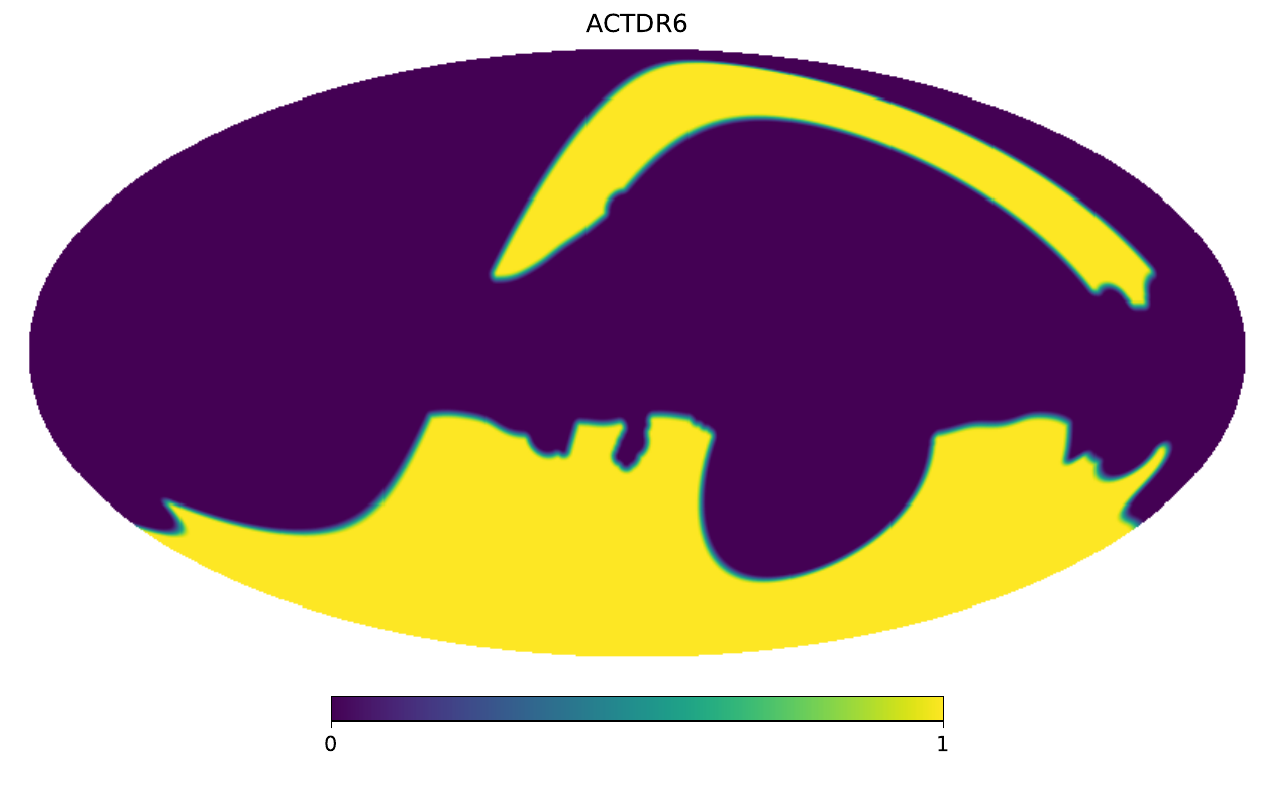}
\caption{
The Planck PR3 and ACTDR6 masks used for the lensing analysis, shown in Galactic coordinates. For the Planck mask, we apply a $0.5$ degree apodization to the public lensing mask.
}
\label{fig:mask}
\ec
\end{figure*}

\begin{figure*}
\bc
\includegraphics[width=8.5cm,clip]{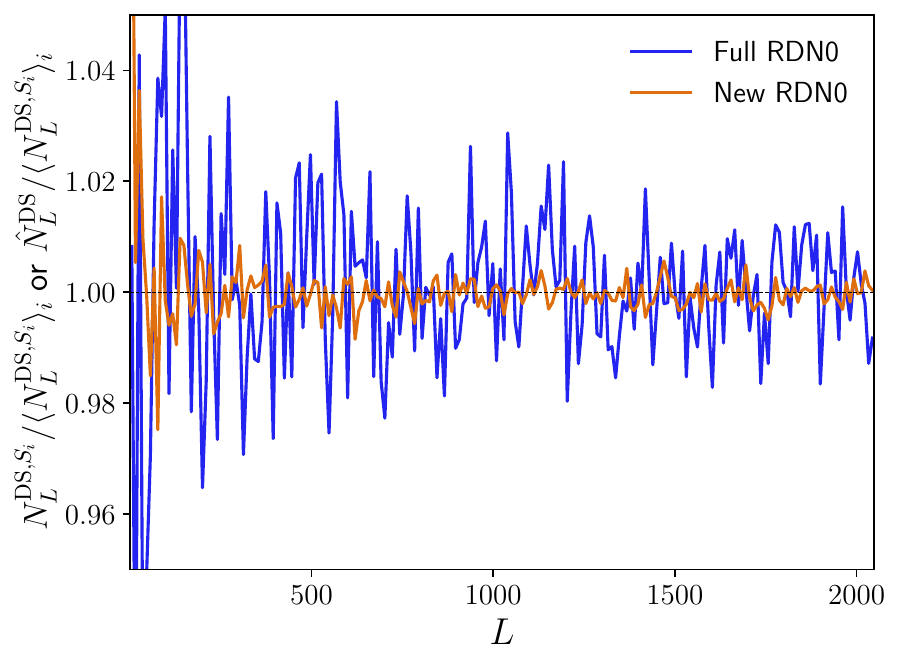}
\includegraphics[width=8.5cm,clip]{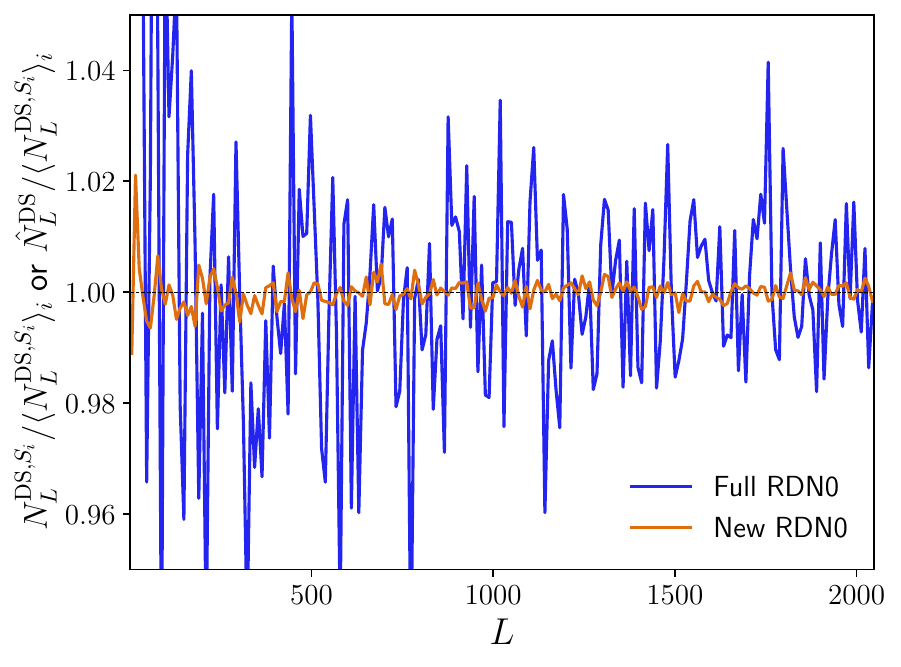}
\caption{
Comparison of the control-variate estimator for the data--simulation cross term in Eq.~\eqref{Eq:RDN0-new:DS} with the conventional full-RDN0 estimator, for the Planck-like case (left) and the ACTDR6-like case (right). Orange and bluse curves show the control-variate and conventional estimates, respectively. For illustration, we use only one MC realization to compute the estimator to see the scatter, while we use many MC realizations to converge the calculation in practice. The results are normalized by $\hat{N}^{\rm DS}_L$ evaluated by averaging over $500$ MC realizations.
}
\label{fig:new-RDN0}
\ec
\end{figure*}

\subsection{A control-variate method}

Here, we propose a new method for accelerating the RDN0 calculation using a control-variate technique. As shown below, this technique relies on constructing a random variable that is highly correlated with the RDN0 estimator, but whose expectation value can be evaluated much more efficiently.

The basic principle of the control-variate method is as follows. Suppose that we wish to estimate a quantity $\theta$ by averaging over many MC realizations, $\theta^i$, with expectation value $\theta$. We consider a situation in which $\theta^i$ has large MC scatter and therefore converges slowly. To reduce this scatter, we introduce a control variate, $\tilde{\theta}^i$, that is highly correlated with $\theta^i$. We can then construct an estimator of $\theta$ with reduced MC scatter as \cite{KahnMarshall:1953,LavenbergWelch:1981,Chartier:2020:carpool}
\al{
    \hat{\theta} = \ave{\theta^i-\alpha\tilde{\theta}^i}_i + \alpha\tilde{\theta}
    \,, \label{Eq:CV}
}
where $\tilde{\theta}$ is the expectation value of $\tilde{\theta}^i$. The coefficient $\alpha$ is chosen to minimize the variance of the estimator, yielding, e.g., \cite{Chartier:2020:carpool}
\al{
    \alpha = \frac{{\rm Cov}(\theta^i,\tilde{\theta}^i)}{\sigma^2(\tilde{\theta}^i)} = \rho(\theta^i,\tilde{\theta}^i) \frac{\sigma(\theta^i)}{\sigma(\tilde{\theta}^i)}
    \,. \label{Eq:alpha}
}
Here, $\sigma^2$ denotes the variance and $\rho$ is the correlation coefficient. With this choice, the variance of the MC average is reduced by a factor $1-\rho^2$. The estimator in Eq.~\eqref{Eq:CV} is unbiased by construction. If the original estimator $\theta^i$ and the control variate $\tilde{\theta}^i$ are strongly correlated, the MC fluctuations in the two terms of Eq.~\eqref{Eq:CV} cancel efficiently. The expectation value $\tilde{\theta}$ must be known analytically, or be computable with negligible computational cost and MC scatter.

We now apply this control-variate construction to the estimators of the data--simulation cross term and the MCN0 term in RDN0. We define
\al{
    \hat{N}^{\rm DS}_L &\equiv \ave{N_L^{{\rm DS},S_i} - \alpha^{\rm DS}_L \tilde{N}_L^{{\rm DS},i}}_i + \alpha^{\rm DS}_L \tilde{N}^{\rm DS}_L
    \,, \label{Eq:RDN0-new:DS} \\ 
    \hat{N}^{\rm MC}_L &\equiv \ave{N_L^{{\rm MC},S_iS'_i} - \alpha^{\rm MC}_L \tilde{N}_L^{{\rm MC},i}}_i + \alpha^{\rm MC}_L \tilde{N}^{\rm MC}_L
    \,, \label{Eq:RDN0-new:MC}
}
where $\tilde{N}_L^{{\rm DS},i}$ and $\tilde{N}_L^{{\rm MC},i}$ are the control variates, and $\alpha^{\rm DS}_L$ and $\alpha^{\rm MC}_L$ are coefficients obtained from Eq.~\eqref{Eq:alpha}. These estimators are unbiased, provided that the expectation values of the control variates, $\tilde{N}^{\rm DS}_L$ and $\tilde{N}^{\rm MC}_L$, are evaluated consistently. The control variates should be highly correlated with the original estimators, while their expectation values should be computable with negligible cost.

A simple possible choice for the control variate is the diagonal approximation to $N_L^{{\rm DS},S_i}$ and $N_L^{{\rm MC},S_iS'_i}$. Although this approximation captures part of the mean bias, it does not track the realization-to-realization fluctuations of the full estimators sufficiently well, because those fluctuations are dominated by off-diagonal covariance of the temperature anisotropies.

To reduce the fluctuations associated with these off-diagonal components, we instead construct control variates using simplified simulations for which the analytic expectation values are known. In this paper, we use statistically isotropic simulations, denoted by $I$, for this purpose. Specifically, we use the input full-sky lensed CMB realizations before applying the survey mask as the statistically isotropic simulations. For the data--simulation cross term, we replace the simulation realization $S_i$ in Eq.~\eqref{Eq:RDN0:DS} with $I_i$:
\al{
    \tilde{N}_L^{{\rm DS},i} = N_L^{{\rm DS},I_i}
    \,. 
}
For the MCN0, we replace one of the simulation realizations $S'_i$ to $I_i$ in Eq.~\eqref{Eq:RDN0:MC}:
\al{
    \tilde{N}_L^{{\rm MC},i} = N_L^{{\rm MC},S_iI_i} 
    \,. 
}
In this case, the expectation values become:
\al{
    \tilde{N}^{\rm DS}_L &= \frac{A_L^2}{2L+1} \sum_{\l\l'}\frac{f^2_{\l L\l'}}{2}\left(\widehat{C}^{\oT\oT}_{\l}C^{\oT\oT}_{\l'}+(\l\leftrightarrow\l')\right)
    \,, \label{Eq:RDN0-new:DS:expect} \\
    \tilde{N}^{\rm MC}_L &= \frac{A_L^2}{2L+1} \sum_{\l\l'}\frac{f^2_{\l L\l'}}{2}\left(C^{\oT\oT,S}_{\l}C^{\oT\oT}_{\l'}+(\l\leftrightarrow\l')\right)
    \,. 
}
Here, $C^{\oT\oT,S}_{\l}=\ave{\sum_m|\oT^{S_i}_{\l m}|^2/(2\l+1)}_i$ is the power spectrum of $\oT$ estimated from the full simulation set $S$. The above expectation values can be computed very efficiently. Importantly, deriving these analytic expectation values does not require assuming that the observed temperature fluctuations at different multipoles are uncorrelated.

One might expect the survey mask to significantly reduce the correlation between the original estimator and the control variate. Naively, the former is sensitive only to fluctuations inside the survey region, whereas the latter is computed from a full-sky statistically isotropic map and therefore contains fluctuations outside the mask. In practice, however, the control variate is also dominated by fluctuations inside the survey region. This is because the lensing reconstruction in the data--simulation cross term correlates two temperature fields, one of which is masked. Since the reconstruction is approximately local, masking one field suppresses contributions from the other field outside the survey region. The reconstructed lensing map is therefore nearly insensitive to fluctuations outside the mask. We explore this more quantitatively in the next subsection.

\subsection{Numerical simulation}

To quantify the correlation between the original estimator and the isotropic control variate, we compute both quantities using simulations for the Planck-like, ACTDR6-like, and SO-like experimental configurations defined below. 
The correlation coefficient between the original estimator and the control variate is similar for the data--simulation cross term and for MCN0. The reason is that the data--simulation term is evaluated at fixed data, whereas MCN0 effectively averages over realizations that play the role of the data field. If the realistic simulations are representative of the data, the correlation structure of the two estimators is therefore similar. In the following, we focus on the data--simulation cross term.

We use two analysis masks: the Planck PR3 lensing analysis mask \cite{PR3:phi} for the Planck-like case, and the ACTDR6 lensing survey mask \cite{Qu:ACT:2023} for the ACTDR6-like and SO-like cases. These masks are shown in Fig.~\ref{fig:mask} in Galactic coordinates. For the Planck mask, we apply a $0.5\,{\rm deg}$ apodization. The ACTDR6 mask is the apodized survey mask and does not include a point-source mask, since inpainting is performed after point-source removal.

We assume that data-split maps are used for the lensing analysis, so that the disconnected bias contribution relevant for RDN0 can be evaluated using signal-only simulations \cite{Madhavacheril:2020:split,Qu:ACT:2023}. We therefore use only realizations of the lensed CMB signal. For the conventional RDN0 estimate, the lensed CMB simulations are multiplied by the corresponding analysis mask, and one realization from this simulation set is treated as the data. For the control variate, we use full-sky lensed CMB simulations without applying the mask.

For the Planck-like case, we use CMB multipoles in the range $100\leq \ell \leq 2048$ for the lensing reconstruction and assume a white-noise level of $50\,\mu{\rm K}$-arcmin with a beam size of $7$ arcmin in the observed temperature power spectrum used for diagonal filtering in the lensing estimator. For the ACTDR6-like case, we use $600\leq \ell \leq 3000$ and assume a noise level of $10\,\mu{\rm K}$-arcmin with a beam size of $1$ arcmin. For the SO-like case, we use the same multipole range as in the ACTDR6-like case, but assume an improved noise level of $4\,\mu{\rm K}$-arcmin. Although we do not include noise realizations in the simulations, the assumed observational noise level enters the diagonal filtering in the lensing estimator of Eq.~\eqref{Eq:estg}. With the above CMB multipole ranges, the temperature multipoles used for the reconstruction are dominated by the CMB signal.

Figure~\ref{fig:new-RDN0} compares the control-variate method of Eq.~\eqref{Eq:RDN0-new:DS} with the conventional approach of Eq.~\eqref{Eq:RDN0:DS} for a fixed realization of the observed temperature anisotropies, $\oT_{\ell m}$, and a simulation realization, $\oT^{S_i}_{\ell m}$. For the control-variate estimator, we choose $\alpha_L=W_6/W_4$, where $W_n=\Int{}{\Omega}{}W^n(\Omega)$ and $W(\Omega)$ denotes the analysis mask in pixel space.
\footnote{This empirical factor provides a simple but effective approximation to the optimal coefficient. This choice is motivated by the fact that the numerator and denominator in Eq.~\eqref{Eq:alpha} contain six and four factors of $W(\Omega)$, respectively. The approximate modification to $\alpha_L$ due to the mask is therefore $W_6/W_4$; see, e.g., \cite{Namikawa:2012:bhe}. Assuming that only the amplitudes of the original estimator and control variate are modified by the mask, a natural choice for $\alpha_L$ in Eq.~\eqref{Eq:RDN0-new:DS} is $W_6/W_4$.}
In both methods, the results are normalized by the analytic expectation given in Eq.~\eqref{Eq:RDN0-new:DS:expect}. Results are shown for the Planck-like and ACTDR6-like cases. The scatter in the control-variate method is significantly suppressed relative to that in the conventional estimator.

\begin{figure*}
\bc
\includegraphics[width=8.8cm,clip]{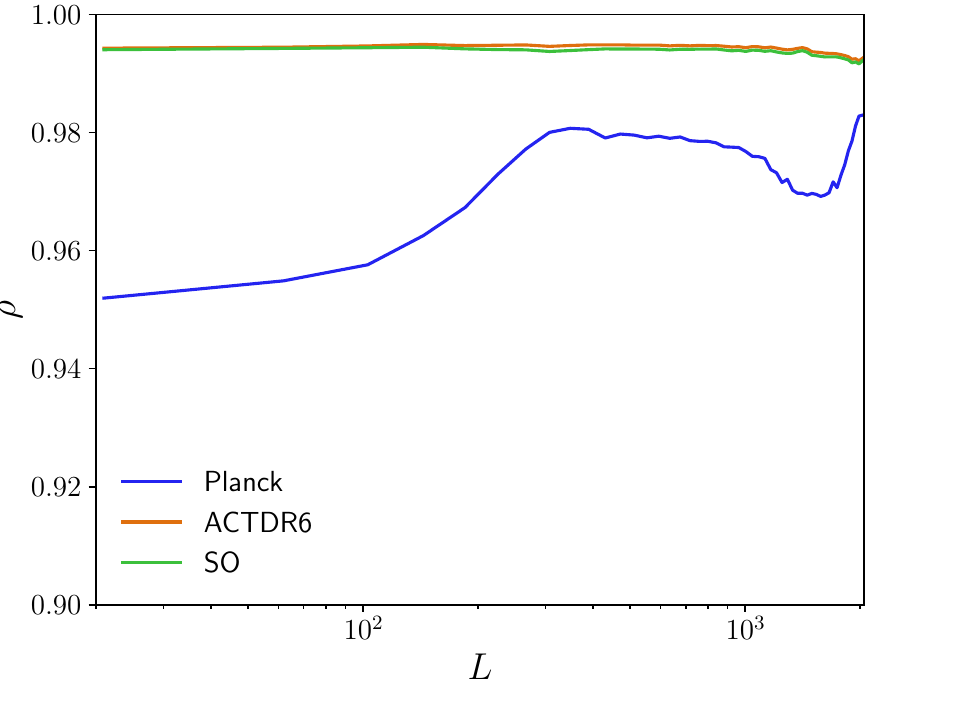}
\includegraphics[width=8.8cm,clip]{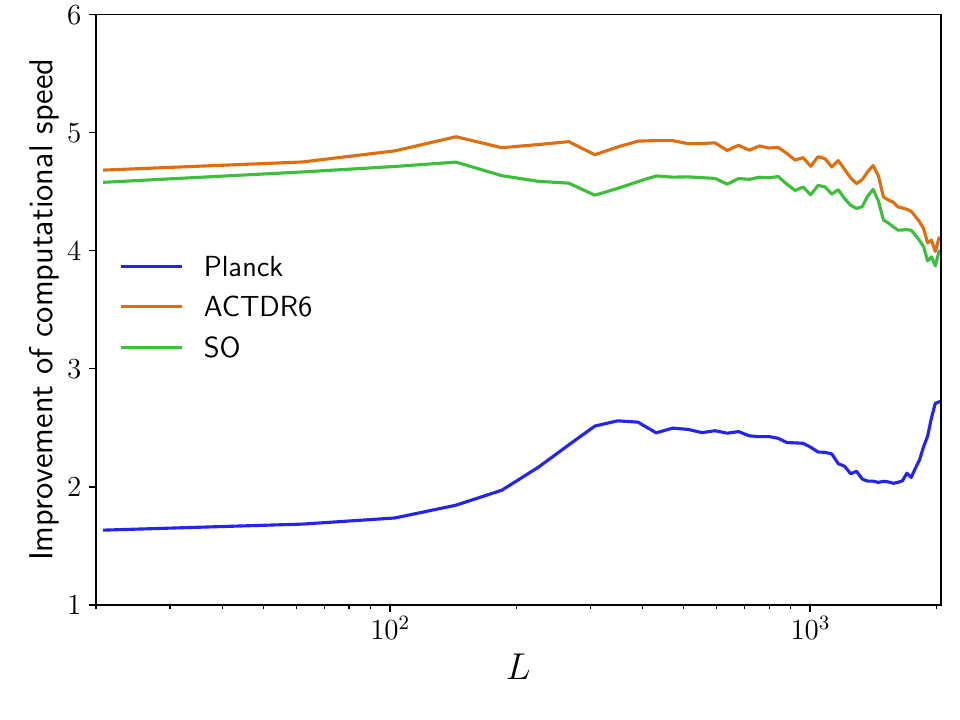}
\caption{
{\it Left}: Correlation coefficients between the original estimator, $N_L^{{\rm DS},S_i}$, and the control variate, $\tilde{N}_L^{{\rm DS},i}=N_L^{{\rm DS},I_i}$, for the Planck-like, ACTDR6-like, and SO-like cases.
{\it Right}: Estimated speed-up factor of the control-variate RDN0 calculation, defined as $1/\beta_L=0.5/\sqrt{1-\rho_L^2}$, as a function of $L$ for the Planck-like, ACTDR6-like, and SO-like cases.
}
\label{fig:correlation:speed}
\ec
\end{figure*}
%

Figure~\ref{fig:correlation:speed} shows the correlation coefficient between the conventional estimator and the control variate. The correlation coefficient is approximately $97\%$ for the Planck-like case and exceeds $99\%$ for the ACTDR6-like and SO-like cases. The correlation coefficients for the ACTDR6-like and SO-like cases are very similar.
Figure~\ref{fig:correlation:speed} also shows the corresponding improvement factor in computational speed. To express the gain in terms of computational time, we define
\al{
    \beta_L \equiv 2\sqrt{1-\rho_L^2}
    \,, \label{Eq:speed}
}
where the prefactor of $2$ accounts for the two RDN0-like evaluations required by the control-variate estimator. The corresponding speed-up factor is
\al{
    \frac{1}{\beta_L} = \frac{0.5}{\sqrt{1-\rho_L^2}}
    \,.
}
The results indicate that the RDN0 computation can be accelerated by a factor of approximately two for the Planck-like case and by a factor of approximately five for the ACTDR6-like and SO-like cases.

\section{Discussion}

%
\begin{figure}
\bc
\includegraphics[width=8.5cm,clip]{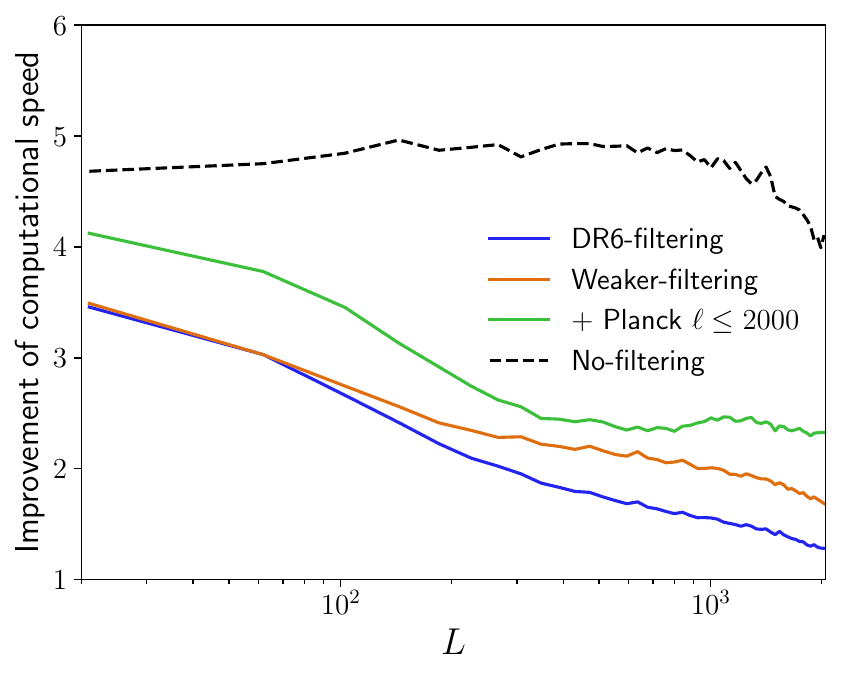}
\caption{
Same as the left panel of Fig.~\ref{fig:correlation:speed}, but for the ACTDR6-like case with Fourier-mode filtering applied. The ACTDR6 filtering removes modes with $|\ell_x|<90$ and $|\ell_y|<50$. We also show cases with weaker filtering, $|\ell_x|<45$ and $|\ell_y|<25$, and with Planck-like information used to fill the missing modes up to $\ell\leq 2000$.
}
\label{fig:filter}
\ec
\end{figure}

In the calculations above, we included no practical effects in the signal simulations other than the analysis mask. Another important effect in realistic analyses is signal filtering. In the ACTDR6 analysis, atmospheric noise and other sources of contamination are removed by applying two-dimensional Fourier-mode filtering to the observed maps. If this filtering is included in the full simulations, it removes part of the CMB signal and therefore reduces the correlation between the conventional RDN0 estimator and the isotropic control variate.

Figure~\ref{fig:filter} shows the scale-dependent correlation coefficient obtained when Fourier-mode filtering is included for the ACTDR6-like case. Here, we remove Fourier modes with $|\ell_x|<90$ and $|\ell_y|<50$. In this baseline filtering case, the correlation is reduced to approximately $94\%$, depending on scale. This corresponds to an approximately $50\%$ improvement in computational speed. We also show a less aggressive filtering case, in which only modes with $|\ell_x|<45$ and $|\ell_y|<25$ are removed. In this case, the correlation improves to approximately $96\%$, corresponding to a speed-up by a factor of about two relative to the conventional RDN0 calculation. However, with this weaker filtering, the lensing measurement may not pass the required null tests. There is therefore a trade-off between computational cost and robustness of the measurement.

One possible way to mitigate the impact of filtering is to construct the control variate using the flat-sky approximation. In the flat-sky limit, even in the presence of Fourier-mode filtering, the expectation value of the isotropic control variate has a numerically tractable analytic expression; see Appendix~\ref{app:flatsky} for details. However, this approach requires Gaussian random fields that are both nearly perfectly correlated with the curved-sky fluctuations and have a covariance that is diagonal in Fourier space. A naive projection of the curved-sky fluctuations onto a two-dimensional grid reduces the correlation too much to provide a useful speed-up of the RDN0 calculation. We considered dividing the survey region into sub-patches, so that the flat-sky approximation is more accurate within each patch, but we were unable to generate random fluctuations that remained sufficiently correlated with the original full-sky realizations. Moreover, this approach requires computing the flat-sky RDN0 for all patches, so the flat-sky calculation must be substantially faster than the full curved-sky RDN0 calculation to be beneficial.

Another possible, and potentially more effective, way to improve the correlation is to fill the missing CMB modes using data from full-sky CMB observations, such as Planck for ongoing experiments and LiteBIRD in the future. For example, the full-sky Planck temperature map is signal-dominated up to $\ell\sim 2000$. Figure~\ref{fig:filter} also shows a case in which some of the missing modes are filled. In this calculation, we simply retain the filtered modes for multipoles $\ell\leq 2000$, mimicking the case in which they are supplied by Planck. We find that the correlation is restored and that the computational cost is again significantly reduced, with a speed-up factor of around three at $L\lesssim 300$, where the lensing signal is most important. Further improvements may be explored in future work.

\section{Summary} \label{sec:summary}

We have proposed a new RDN0 algorithm that accelerates the RDN0 calculation for CMB lensing measurements. The method is based on a control-variate construction in which the conventional RDN0 estimator is combined with RDN0-like control variates computed from statistically isotropic lensed CMB simulations. We found that the isotropic control variate is highly correlated with the conventional RDN0 estimator, leading to reductions in computational cost by factors of approximately two and five for Planck-like and ACTDR6-like experiments, respectively. For ACTDR6-like and SO-like configurations, however, this correlation is degraded by Fourier-mode filtering. The resulting improvement in computational speed therefore depends on how many Fourier modes can be retained while still satisfying the required null tests.

Throughout this work, we assumed a lensing measurement based on data-split maps in order to avoid possible noise correlations in the RDN0 estimate. To apply our method to a lensing measurement using the full data set, one would need to model the noise component in the observed CMB maps. Since the noise is generally not statistically isotropic, the simple isotropic-simulation construction used in this paper cannot be applied directly.

For a two-split analysis, such as in Planck, the proposed method can still be used for RDN0 if the lensing power spectrum is estimated by cross-correlating reconstructions from different data splits, for example $\estk[\Theta^A,\Theta^A]$ and $\estk[\Theta^B,\Theta^B]$. The noise may contribute to the mean-field bias of each reconstruction, but, because the noise in different splits is uncorrelated, it does not contribute to the RDN0 term in this cross-spectrum. In space-based experiments such as Planck, LiteBIRD, and PICO \cite{PICO:2019}, the loss of correlation caused by Fourier-mode filtering is not expected to be relevant, since atmospheric noise filtering is not required. Therefore, for these experiments, our method should provide an efficient way to accelerate the RDN0 computation.

We have also neglected foreground contributions in the lensing measurement. For foreground components that can be described as statistically isotropic fields, our algorithm can be applied straightforwardly. To a good approximation, extragalactic foregrounds are expected to be statistically isotropic at the level of the two-point angular power spectrum, meaning that their covariance does not contain off-diagonal correlations, even if the fields themselves are non-Gaussian. We note that, in the recent ACT analysis \cite{Qu:ACT:2023}, foreground contributions were not included in the RDN0 estimation, because their impact is already suppressed by the use of RDN0 and was found to be negligible in the analysis. Galactic foregrounds, on the other hand, are spatially inhomogeneous and may induce off-diagonal covariance. Nevertheless, their contribution is expected to be small on the small angular scales most relevant for CMB lensing reconstruction.

In this work, we used a simple diagonal filter for the inverse-variance filtering of the observed temperature anisotropies. A more optimal analysis would use non-diagonal filtering, which is computationally more expensive. Such filtering is particularly important for polarization data, since survey boundaries mix $E$- and $B$-mode polarization \cite{Lewis:2001:EB,Bunn:2002,Smith:2005:chi-estimator}. The resulting $E$-to-$B$ leakage in the $B$-mode map can make the estimator suboptimal if it is not properly accounted for. In contrast, polarization-based estimators are expected to be much less sensitive to the Fourier-filtering issue discussed above. Our control-variate approach may therefore be especially efficient for polarization-based lensing analyses. A detailed study of the corresponding control variate for polarization is left for future work.

Finally, we note that the proposed RDN0 algorithm is general and can be applied to other four-point statistics with a structure similar to CMB lensing. RDN0 debiasing has recently been used in analyses of anisotropic cosmic birefringence, patchy reionization, lensing curl modes, and other line-of-sight distortions; see, e.g., \cite{Namikawa:2020ffr,Namikawa:2017uke,Namikawa:2011:curlrec,BICEPKeck:2022:los-dist}. The efficiency of our control-variate RDN0 algorithm for these and other four-point analyses can be explored in future work.


\begin{acknowledgments}
We acknowledge support from JSPS KAKENHI Grant No. JP25K00996, and No. JP26H00405, and from JST EXPERT-J, Japan Grant No. JPMJEX2508. The Kavli IPMU is supported by World Premier International Research Center Initiative (WPI Initiative), MEXT, Japan. This work uses resources of the National Energy Research Scientific Computing Center (NERSC). 
\end{acknowledgments}

\section*{Data availability}
The data that support the findings of this article are not publicly available. The data are available from the authors upon reasonable request.

\onecolumngrid

\appendix

\section{Flat-sky Fourier-mode filtering} \label{app:flatsky}

In this appendix, we discuss a possible construction of a control variate in the flat-sky approximation. We focus on the data--simulation cross term.

\subsection{RDN0 in the flat-sky approximation}

We first summarize the expression for RDN0 in the flat-sky approximation. The data--simulation cross term in the flat-sky RDN0 is given by
\al{
    (2\pi)^2 N^{\rm f}_L \equiv (A^{\rm f}_L)^2\Int{2}{\bl}{(2\pi)^2}\Int{2}{\bl'}{(2\pi)^2} f_{\bl,\bL}f_{\bl',\bL}\ave{\oT_{\bl}\oT^*_{\bl'}} \oT_{\bu}\oT^*_{\bu'} + (\text{$3$ perms.})
    \,, \label{Eq:flat:RDN0}
}
where $A^{\rm f}_L$ is the normalization of the flat-sky estimator, $\oT_{\bl}$ denotes the Fourier mode of the inverse-variance-filtered temperature anisotropies, and we have defined \cite{HuOkamoto:2001}
\al{
    f_{\bl,\bL} = \frac{1}{2}[(\bL\cdot\bl)\CTT_\l+(\bL\cdot\bu)\CTT_{\bu}]
    \,, 
}
with $\bu=\bL-\bl$. Equation~\eqref{Eq:flat:RDN0} is the flat-sky counterpart of Eq.~\eqref{Eq:RDN0:DS}. Using MC simulations, this term can be estimated as
\al{
    (2\pi)^2 N^{\rm f}_L = \AVE{(A^{\rm f}_L)^2\Int{2}{\bl}{(2\pi)^2}\Int{2}{\bl'}{(2\pi)^2} f_{\bl,\bL}f_{\bl',\bL}\oT_{\bl}^{S_i}(\oT_{\bl'}^{S_i})^* \oT_{\bu}\oT^*_{\bu'} + (\text{$3$ perms.}) }_i 
    \,. \label{Eq:NSxD:def}
}
This calculation requires averaging over many simulation realizations.

\subsection{A flat-sky control variate for RDN0} 

For ACT-like simulations, we assume that the observed temperature map is first projected onto a CAR pixelization. The analysis mask is then applied, followed by a Fourier transform. We further apply the Fourier-space filter to the temperature modes, transform the filtered map back to real space, and finally compute the spherical-harmonic coefficients $\oT_{\ell m}$ of the filtered map. In the presence of this filtering, the ensemble average of the RDN0 contribution from an isotropic simulation cannot be computed analytically in the curved-sky treatment, because the filtering makes the isotropic input map statistically anisotropic.

In the flat-sky approximation, however, the corresponding ensemble average can still be computed analytically, even in the presence of Fourier-mode filtering. This motivates a control variate based on the flat-sky RDN0. We define
\al{
    \hat{N}^{\rm DS}_L = \ave{N_L^{{\rm DS},S_i}-\alpha_L\tilde{N}_L^{{\rm DS-f},i}}_i + \alpha_L\tilde{N}_L^{\rm DS-f}
    \,. \label{Eq:RDN0-new-f}
}
The first term inside the brackets is the conventional full RDN0 computed on the curved sky, while the second term is the flat-sky RDN0 corresponding to Eq.~\eqref{Eq:NSxD:def}. In the latter term, the observed temperature Fourier modes are obtained by Fourier transforming the temperature map projected onto a two-dimensional grid.

For the simulated temperature Fourier modes $\oT_{\bl}^{S_i}$, two conditions must be satisfied. First, the resulting flat-sky RDN0 must be strongly correlated with the curved-sky RDN0 in the first term of Eq.~\eqref{Eq:RDN0-new-f}. Second, the covariance of the simulated Fourier modes must be diagonal in Fourier space:
\al{
    \ave{\oT_{\bl}^{S_i}(\oT_{\bl'}^{S_i})^*}_i = (2\pi)^2\delta^{\rm D}(\bl-\bl')F^2_{\bl}C^{\oT\oT,S}_\ell
    \,. \label{Eq:flat:diagonal}
}
Here, $\delta^{\rm D}$ is the two-dimensional delta function, $F_{\bl}$ denotes the two-dimensional Fourier-mode filter, and we have assumed that the isotropic temperature Fourier modes are filtered by multiplying them by $F_{\bl}$. Using Eq.~\eqref{Eq:flat:diagonal}, the ensemble average in the last term of Eq.~\eqref{Eq:RDN0-new-f} can be computed with an FFT-based algorithm:
\al{
    \tilde{N}^{\rm DS-f}_L
    &= \frac{(A_L^{\rm f})^2}{2}\Int{2}{\hatn}{}[\pd_i\pd_j\E^{-\iu\bL\cdot\hatn}]
    \bigg[
    \pd^i\pd^j\Int{2}{\bl}{(2\pi)^2}\E^{\iu\bl\cdot\hatn}(\CTT_\l)^2 F^2_{\bl}C^{\oT\oT,S}_\ell 
    \Int{2}{\bu}{(2\pi)^2}\E^{\iu\bu\cdot\hatn}\hC_{\bu}^{\oT\oT}
    \notag \\
    &\qquad+ \Int{2}{\bl}{(2\pi)^2}\pd^i\E^{\iu\bl\cdot\hatn}\CTT_\l F^2_{\bl}C^{\oT\oT,S}_\ell 
    \Int{2}{\bu}{(2\pi)^2}\pd^j\E^{\iu\bu\cdot\hatn}\CTT_\ell \hC_{\bu}^{\oT\oT}
    \bigg]
    \,, 
}
where we define $|\oT_{\bl}|^2=(2\pi)^2\delta^{\rm D}(\bm{0})\hC_{\bl}^{\oT\oT}$.

The main difficulty with this flat-sky control variate is generating random fluctuations that satisfy the first condition above. We attempted to generate random fields on a two-dimensional grid that are highly correlated with the original full-sky realizations. For example, we considered dividing the survey region into multiple subpatches. In each subpatch, we projected the full-sky map onto a two-dimensional grid and computed the corresponding Fourier modes. We then evaluated the correlation coefficient between the angular power spectrum estimated from these two-dimensional fluctuations and that obtained from the full-sky spherical-harmonic coefficients. However, with this approach, we were unable to obtain fluctuations with correlations greater than $90\%$. This level of correlation is insufficient to provide a useful speed-up of the RDN0 calculation.

\bibliographystyle{mybst}
\bibliography{cite}

@article{Ding2022DESICARPool,
  author        = {Ding, Zhejie and Chuang, Chia-Hsun and Yu, Yu and Garrison, Lehman H. and Bayer, Adrian E. and Feng, Yu and Modi, Chirag and Eisenstein, Daniel J. and White, Martin and Variu, Andrei and Zhao, Cheng and Zhang, Hanyu and Meneses Rizo, Jennifer and Brooks, David and Dawson, Kyle and Doel, Peter and Gaztanaga, Enrique and Kehoe, Robert and Krolewski, Alex and Landriau, Martin and Palanque-Delabrouille, Nathalie and Poppett, Claire},
  title         = {The {DESI} {$N$}-body Simulation Project -- {II}. Suppressing sample variance with fast simulations},
  journal       = {\mnras},
  volume        = {514},
  number        = {3},
  pages         = {3308--3326},
  year          = {2022},
  doi           = {10.1093/mnras/stac1501},
  eprint        = {2202.06074},
  archivePrefix = {arXiv},
  primaryClass  = {astro-ph.CO}
}

@article{DeRose2023ZeldovichControlVariatesRSD,
  author        = {DeRose, Joseph and Chen, Shi-Fan and Kokron, Nickolas and White, Martin},
  title         = {Precision redshift-space galaxy power spectra using {Zel'dovich} control variates},
  journal       = {\jcap},
  volume        = {2023},
  number        = {02},
  pages         = {008},
  year          = {2023},
  doi           = {10.1088/1475-7516/2023/02/008},
  eprint        = {2210.14239},
  archivePrefix = {arXiv},
  primaryClass  = {astro-ph.CO}
}

@article{Hadzhiyska2023AnalyticControlVariatesDESI,
  author        = {Hadzhiyska, Boryana and White, Martin J. and Chen, Xinyi and Garrison, Lehman H. and DeRose, Joseph and Padmanabhan, Nikhil and Garcia-Quintero, Cristhian and Mena-Fern{\'a}ndez, Juan and Chen, Shi-Fan and Seo, Hee-Jong and McDonald, Patrick and Aguilar, Jessica and Ahlen, Steven and Brooks, David and Claybaugh, Todd and de la Macorra, Axel and Doel, Peter and Font-Ribera, Andreu and Forero-Romero, Jaime E. and Gontcho A Gontcho, Satya and Honscheid, Klaus and Kremin, Anthony and Landriau, Martin and Manera, Marc and Miquel, Ramon and Nie, Jundan and Palanque-Delabrouille, Nathalie and Rezaie, Mehdi and Rossi, Graziano and Sanchez, Eusebio and Schubnell, Michael and Tarl{\'e}, Gregory and Zhou, Zhimin},
  title         = {Mitigating the noise of {DESI} mocks using analytic control variates},
  journal       = {The Open Journal of Astrophysics},
  volume        = {6},
  pages         = {23},
  year          = {2023},
  doi           = {10.21105/astro.2308.12343},
  eprint        = {2308.12343},
  archivePrefix = {arXiv},
  primaryClass  = {astro-ph.CO}
}

@article{Doytcheva2024HydroControlVariates,
  author        = {Doytcheva, Alexandra and Gerou, Filomela V. and Lange, Johannes U.},
  title         = {High-Precision Galaxy Clustering Predictions from Small-Volume Hydrodynamical Simulations via Control Variates},
  journal       = {arXiv e-prints},
  pages         = {arXiv:2410.14546},
  year          = {2024},
  eprint        = {2410.14546},
  archivePrefix = {arXiv},
  primaryClass  = {astro-ph.CO}
}

@article{Hadzhiyska2025LyalphaBAOControlVariates,
  author        = {Hadzhiyska, Boryana and de Belsunce, Roger and Cuceu, Andrei and Guy, Julien and Ivanov, Mikhail M. and Coquinot, Henri and Font-Ribera, Andreu},
  title         = {Measuring and unbiasing the {BAO} shift in the {Lyman-Alpha} forest with {AbacusSummit}},
  journal       = {\mnras},
  volume        = {540},
  number        = {2},
  pages         = {1960--1978},
  year          = {2025},
  doi           = {10.1093/mnras/staf824},
  eprint        = {2503.13442},
  archivePrefix = {arXiv},
  primaryClass  = {astro-ph.CO}
}

@article{Kokron2025BispectrumControlVariates,
  author        = {Kokron, Nickolas and Chen, Shi-Fan},
  title         = {Control variates from {Eulerian} and {Lagrangian} perturbation theory: Application to the bispectrum},
  journal       = {arXiv e-prints},
  pages         = {arXiv:2510.07375},
  year          = {2025},
  eprint        = {2510.07375},
  archivePrefix = {arXiv},
  primaryClass  = {astro-ph.CO}
}

@inbook{Bianchini:2025:review,
 author = "Bianchini, Federico and Maniyar, Abhishek S.",
 title = "The Encyclopedia of Astrophysics",
 chapter = "{The Cosmic Microwave Background -- Secondary Anisotropies}",
 eprint = "2501.13913",
 month = "1",
 publisher = "Elsevier",
 year = "2025"
}

@article{Blanchard:1987AA,
 author = {Blanchard, A. and Schneider, J.},
 journal = {\aap},
 pages = {1-6},
 title = {Gravitational lensing effect on the fluctuations of the cosmic background radiation},
 volume = {184},
 year  = {1987},
}

@article{Bunn:2002,
 author = "Bunn, Emory F. and Zaldarriaga, Matias and Tegmark, Max and de Oliveira-Costa, Angelica",
 title = "{E/B decomposition of finite pixelized CMB maps}",
 eprint = "astro-ph/0207338",
 doi = "10.1103/PhysRevD.67.023501",
 journal = "\prd",
 volume = "67",
 pages = "023501",
 year = "2003"
}

@article{Chartier:2020:carpool,
    author = "Chartier, Nicolas and Wandelt, Benjamin and Akrami, Yashar and Villaescusa-Navarro, Francisco",
    title = "{CARPool: fast, accurate computation of large-scale structure statistics by pairing costly and cheap cosmological simulations}",
    eprint = "2009.08970",
    archivePrefix = "arXiv",
    primaryClass = "astro-ph.CO",
    doi = "10.1093/mnras/stab430",
    journal = "\mnras",
    volume = "503",
    number = "2",
    pages = "1897--1914",
    year = "2021"
}

@article{Chartier:2022:carpool,
    author = "Chartier, Nicolas and Wandelt, Benjamin D.",
    title = "{Bayesian control variates for optimal covariance estimation with pairs of simulations and surrogates}",
    eprint = "2204.03070",
    archivePrefix = "arXiv",
    primaryClass = "astro-ph.CO",
    doi = "10.1093/mnras/stac1837",
    journal = "\mnras",
    volume = "515",
    number = "1",
    pages = "1296--1315",
    year = "2022"
}

@article{Chudaykin:2025gdn,
    author = "Chudaykin, Anton and Kunz, Martin and Carron, Julien",
    title = "{Modified gravity constraints with the Planck ISW-lensing bispectrum}",
    eprint = "2503.09893",
    archivePrefix = "arXiv",
    primaryClass = "astro-ph.CO",
    doi = "10.1103/9zjh-8htv",
    journal = "\prd",
    volume = "112",
    number = "8",
    pages = "083537",
    year = "2025"
}

@article{PICO:2019,
    author = "Hanany, Shaul and others",
    collaboration = "NASA PICO",
    title = "{PICO: Probe of Inflation and Cosmic Origins}",
    eprint = "1902.10541",
    archivePrefix = "arXiv",
    primaryClass = "astro-ph.IM",
    month = "3",
    year = "2019"
}

@article{Hanson:2009:noise,
 author = {Hanson, D. and Rocha, G. and Gorski, K.},
 eprint = {0907.1927},
 journal = {\mnras},
 pages = {2169-2173},
 title = {Lensing reconstruction from PLANCK sky maps: inhomogeneous noise},
 volume = {400},
 year  = {2009},
}

@article{Hanson:2009:review,
 author = {Hanson, D. and Challinor, A. and Lewis, A.},
 eprint = {0911.0612},
 journal = {Gen. Rel. Grav.},
 pages = {2197-2218},
 title = {Weak lensing of the CMB},
 volume = {42},
 year  = {2010},
}

@article{Hanson:2010:N2,
 author = {Hanson, D. and Challinor, A. and Efstathiou, G. and Bielewicz, P.},
 eprint = {1008.4403},
 journal = {\prd},
 pages = {043005},
 title = {CMB temperature lensing power reconstruction},
 volume = {83},
 year  = {2011},
}

@article{Hu:2001:cmbrecon,
 author = {Hu, W.},
 eprint = {astro-ph/0105424},
 journal = {\apj},
 pages = {L79-L83},
 title = {Mapping the dark matter through the cosmic microwave background damping tail},
 volume = {557},
 year  = {2001},
}

@article{HuOkamoto:2001,
 author = {Hu, W. and Okamoto, T.},
 eprint = {astro-ph/0111606},
 journal = {\apj},
 pages = {566-574},
 title = {Mass Reconstruction with CMB Polarization},
 volume = {574},
 year  = {2002},
}

@article{Jhaveri:2025:tau,
    author = "Jhaveri, Tanisha and Karwal, Tanvi and Hu, Wayne",
    title = "{Turning a negative neutrino mass into a positive optical depth}",
    eprint = "2504.21813",
    archivePrefix = "arXiv",
    primaryClass = "astro-ph.CO",
    month = "4",
    year = "2025"
}

@article{KahnMarshall:1953,
  author       = {H. Kahn and
                  A. W. Marshall},
  title        = {Methods of Reducing Sample Size in Monte Carlo Computations},
  journal      = {Oper. Res.},
  volume       = {1},
  number       = {5},
  pages        = {263--278},
  year         = {1953},
  url          = {https://doi.org/10.1287/opre.1.5.263},
  doi          = {10.1287/OPRE.1.5.263},
  bibsource    = {dblp computer science bibliography, https://dblp.org}
}

@article{Kokron:2022,
    author = "Kokron, Nickolas and Chen, Shi-Fan and White, Martin and DeRose, Joseph and Maus, Mark",
    title = "{Accurate predictions from small boxes: variance suppression via the Zel'dovich approximation}",
    eprint = "2205.15327",
    doi = "10.1088/1475-7516/2022/09/059",
    journal = "\jcap",
    volume = "09",
    pages = "059",
    year = "2022"
}

@article{LavenbergWelch:1981,
 ISSN = {00251909, 15265501},
 URL = {http://www.jstor.org/stable/2631207},
 author = {S. S. Lavenberg and P. D. Welch},
 journal = {Management Science},
 number = {3},
 pages = {322--335},
 publisher = {INFORMS},
 title = {A Perspective on the Use of Control Variables to Increase the Efficiency of Monte Carlo Simulations},
 urldate = {2026-05-15},
 volume = {27},
 year = {1981}
}

@article{Lewis:2001:EB,
 author = "Lewis, Antony and Challinor, Anthony and Turok, Neil",
 title = "{Analysis of CMB polarization on an incomplete sky}",
 eprint = "astro-ph/0106536",
 doi = "10.1103/PhysRevD.65.023505",
 journal = "\prd",
 volume = "65",
 pages = "023505",
 year = "2002"
}

@article{Lewis:2006:review,
 author = {Lewis, A. and Challinor, A.},
 eprint = {astro-ph/0601594},
 journal = {Phys. Rep.},
 pages = {1-65},
 title = {Weak gravitational lensing of the CMB},
 volume = {429},
 year  = {2006},
}

@article{Lewis:2011:bispec,
 author = {Lewis, A. and Challinor, A. and Hanson, D.},
 eprint = {1101.2234},
 journal = {\jcap},
 pages = {018},
 title = {The shape of the CMB lensing bispectrum},
 volume = {03},
 year  = {2011},
}

@article{LiteBIRD:2023phi,
    author = "Lonappan, A. I. and others",
    collaboration = "LiteBIRD",
    title = "{LiteBIRD science goals and forecasts: a full-sky measurement of gravitational lensing of the CMB}",
    eprint = "2312.05184",
    archivePrefix = "arXiv",
    primaryClass = "astro-ph.CO",
    doi = "10.1088/1475-7516/2024/06/009",
    journal = "\jcap",
    volume = "06",
    pages = "009",
    year = "2024"
}

@article{LiteBIRD:2025:phi,
    author = "Ruiz-Granda, M. and others",
    collaboration = "LiteBIRD",
    title = "{LiteBIRD science goals and forecasts: improved full-sky reconstruction of the gravitational lensing potential through the combination of Planck and LiteBIRD data}",
    eprint = "2507.22618",
    archivePrefix = "arXiv",
    primaryClass = "astro-ph.CO",
    doi = "10.1088/1475-7516/2025/11/073",
    journal = "\jcap",
    volume = "11",
    pages = "073",
    year = "2025"
}

@article{Madhavacheril:2020:split,
 author = "Madhavacheril, Mathew S. and Smith, Kendrick M. and Sherwin, Blake D. and Naess, Sigurd",
 title = "{CMB lensing power spectrum estimation without instrument noise bias}",
 eprint = "2011.02475",
 doi = "10.1088/1475-7516/2021/05/028",
 month = "11",
 journal = "\jcap",
 year = "2021",
 pages = "028",
 volume = "2021",
}

@article{Madhavacheril:2024:review,
 author = "Madhavacheril, Mathew S.",
 title = "{Assessing the growth of structure over cosmic time with cosmic microwave backgroundlensing}",
 eprint = "2411.08152",
 doi = "10.1098/rsta.2024.0025",
 journal = "Phil. Trans. Roy. Soc. Lond. A",
 volume = "383",
 number = "2290",
 pages = "20240025",
 year = "2025"
}

@article{Namikawa:2011:curlrec,
 author = {Namikawa, T. and Yamauchi, D. and Taruya, A.},
 eprint = {1110.1718},
 journal = {\jcap},
 pages = {007},
 title = {Full-sky lensing reconstruction of gradient and curl modes from CMB maps},
 volume = {1201},
 year  = {2012},
}

@article{Namikawa:2012:bhe,
 author = {Namikawa, T. and Hanson, D. and Takahashi, R.},
 doi = {10.1093/mnras/stt195},
 eprint = {1209.0091},
 journal = {\mnras},
 pages = {609-620},
 title = {Bias-Hardened CMB Lensing},
 volume = {431},
 year  = {2013},
}

@article{Namikawa:2017uke,
    author = "Namikawa, Toshiya",
    title = "{Constraints on Patchy Reionization from Planck CMB Temperature Trispectrum}",
    eprint = "1711.00058",
    archivePrefix = "arXiv",
    primaryClass = "astro-ph.CO",
    doi = "10.1103/PhysRevD.97.063505",
    journal = "Phys. Rev. D",
    volume = "97",
    number = "6",
    pages = "063505",
    year = "2018"
}

@article{Namikawa:2020ffr,
    author = "Namikawa, Toshiya and others",
    title = "{Atacama Cosmology Telescope: Constraints on cosmic birefringence}",
    eprint = "2001.10465",
    archivePrefix = "arXiv",
    primaryClass = "astro-ph.CO",
    doi = "10.1103/PhysRevD.101.083527",
    journal = "\prd",
    volume = "101",
    number = "8",
    pages = "083527",
    year = "2020"
}

@article{OkamotoHu:quad,
 author = {Okamoto, T. and Hu, W.},
 eprint = {astro-ph/0301031},
 journal = {\prd},
 pages = {083002},
 title = {CMB Lensing Reconstruction on the Full Sky},
 volume = {67},
 year  = {2003},
}

@article{Omori:2017:SPT+Planck,
    author = "{Omori}, Y. and {Chown}, R. and {Simard}, G. and {Story}, K.~T. and {Aylor}, K. and {Baxter}, E.~J. and {Benson}, B.~A. and {Bleem}, L.~E. and {Carlstrom}, J.~E. and {Chang}, C.~L. and {Cho}, H.-M. and {Crawford}, T.~M. and {Crites}, A.~T. and {de Haan}, T. and {Dobbs}, M.~A. and {Everett}, W.~B. and {George}, E.~M. and {Halverson}, N.~W. and {Harrington}, N.~L. and {Holder}, G.~P. and {Hou}, Z. and {Holzapfel}, W.~L. and {Hrubes}, J.~D. and {Knox}, L. and {Lee}, A.~T. and {Leitch}, E.~M. and {Luong-Van}, D. and {Manzotti}, A. and {Marrone}, D.~P. and {McMahon}, J.~J. and {Meyer}, S.~S. and {Mocanu}, L.~M. and {Mohr}, J.~J. and {Natoli}, T. and {Padin}, S. and {Pryke}, C. and {Reichardt}, C.~L. and {Ruhl}, J.~E. and {Sayre}, J.~T. and {Schaffer}, K.~K. and {Shirokoff}, E. and {Staniszewski}, Z. and {Stark}, A.~A. and {Vanderlinde}, K. and {Vieira}, J.~D. and {Williamson}, R. and {Zahn}, O.",
    title = "{A 2500 deg$^2$ CMB Lensing Map from Combined South Pole Telescope and Planck Data}",
    eprint = "1705.00743",
    reportNumber = "FERMILAB-PUB-17-140-AE",
    doi = "10.3847/1538-4357/aa8d1d",
    journal = "\apj",
    volume = "849",
    number = "2",
    pages = "124",
    year = "2017"
}

@article{Qu:2025:ACT+Planck+SPT,
    author = "Qu, Frank J. and others",
    collaboration = "SPT-3G, ACT",
    title = "{Unified and consistent structure growth measurements from joint ACT, SPT and {\textbackslash}textit{Planck} CMB lensing}",
    eprint = "2504.20038",
    reportNumber = "FERMILAB-PUB-25-0279-PPD",
    month = "4",
    year = "2025"
}

@article{Smith:2005:chi-estimator,
 author = {Smith, Kendrick M.},
 doi = {10.1103/PhysRevD.74.083002},
 eprint = {astro-ph/0511629},
 journal = {\prd},
 pages = {083002},
 title = {Pseudo-c(l) estimators which do not mix E and B modes},
 volume = {74},
 year = {2006},
}

@ARTICLE{Das:2011,
 author = {Das, S. and Sherwin, B. D. and Aguirre, P. and Appel, J. W. and Bond, J. R. and Carvalho, C. S. and Devlin, M. J. and Dunkley, J. and D{\"u}nner, R. and {Essinger-Hileman}, T. and Fowler, J. W. and Hajian, A. and Halpern, M. and Hasselfield, M. and Hincks, A. D. and Hlozek, R. and Huffenberger, K. M. and Hughes, J. P. and Irwin, K. D. and Klein, J. and Kosowsky, A. and Lupton, R. H. and Marriage, T. A. and Marsden, D. and Menanteau, F. and Moodley, K. and Niemack, M. D. and Nolta, M. R. and Page, L. A. and Parker, L. and Reese, E. D. and Schmitt, B. L. and Sehgal, N. and Sievers, J. and Spergel, D. N. and Staggs, S. T. and Swetz, D. S. and Switzer, E. R. and Thornton, R. and Visnjic, K. and Wollack, E.},
 collaboration = {ACT},
 title = "{Detection of the Power Spectrum of Cosmic Microwave Background Lensing by the Atacama Cosmology Telescope}",
 journal = {\prl},
 year = 2011,
 volume = {107},
 number = {2},
 eid = {021301},
 pages = {021301},
 doi = {10.1103/PhysRevLett.107.021301},
 eprint = {1103.2124},
}

@article{Sherwin:2011:DE,
 author = {Sherwin, B. D. and others},
 collaboration = {ACT},
 eprint = {1105.0419},
 journal= {\prl},
 pages  = {021302},
 title  = {Evidence for dark energy from the cosmic microwave background alone using the Atacama Cosmology Telescope lensing measurements},
 volume = {107},
 year   = {2011},
}

@article{ACT16:phi,
 author = {{Sherwin}, B. D. and {van Engelen}, A. and {Sehgal}, N. and {Madhavacheril}, M. and {Addison}, G. E. and {Aiola}, S. and {Allison}, R. and {Battaglia}, N. and {Becker}, D. T. and {Beall}, J. A. and {Bond}, J. R. and {Calabrese}, E. and {Datta}, R. and {Devlin}, M. J. and {D{\"u}nner}, R. and {Dunkley}, J. and {Fox}, A. E. and {Gallardo}, P. and {Halpern}, M. and {Hasselfield}, M. and {Henderson}, S. and {Hill}, J. C. and {Hilton}, G. C. and {Hubmayr}, J. and {Hughes}, J. P. and {Hincks}, A. D. and {Hlozek}, R. and {Huffenberger}, K. M. and {Koopman}, B. and {Kosowsky}, A. and {Louis}, T. and {Maurin}, L. and {McMahon}, J. and {Moodley}, K. and {Naess}, S. and {Nati}, F. and {Newburgh}, L. and {Niemack}, M. D. and {Page}, L. A. and {Sievers}, J. and {Spergel}, D. N. and {Staggs}, S. T. and {Thornton}, R. J. and {Van Lanen}, J. and {Vavagiakis}, E. and {Wollack}, E. J.},
 eprint = {1611.09753},
 collaboration = {ACT},
 title = "{The Atacama Cosmology Telescope: Two-Season ACTPol Lensing Power Spectrum}",
 year = {2017},
 journal = {\prd},
 volume = {95},
 pages = {123529},
 doi = {10.1103/PhysRevD.95.123529},
}

@article{MacCrann:ACT:2023,
    author = "MacCrann, Niall and others",
    collaboration = "ACT",
    title = "{The Atacama Cosmology Telescope: Mitigating the Impact of Extragalactic Foregrounds for the DR6 Cosmic Microwave Background Lensing Analysis}",
    eprint = "2304.05196",
    reportNumber = "FERMILAB-PUB-23-236-PPD",
    doi = "10.3847/1538-4357/ad2610",
    journal = "\apj",
    volume = "966",
    number = "1",
    pages = "138",
    year = "2024"
}

@article{Madhavacheril:ACT:2023,
    author = "Madhavacheril, Mathew S. and others",
    collaboration = "ACT",
    title = "{The Atacama Cosmology Telescope: DR6 Gravitational Lensing Map and Cosmological Parameters}",
    eprint = "2304.05203",
    reportNumber = "FERMILAB-PUB-23-206-PPD",
    doi = "10.3847/1538-4357/acff5f",
    journal = "\apj",
    volume = "962",
    number = "2",
    pages = "113",
    year = "2024"
}

@article{Qu:ACT:2023,
    author = "Qu, Frank J. and others",
    collaboration = "ACT",
    title = "{The Atacama Cosmology Telescope: A Measurement of the DR6 CMB Lensing Power Spectrum and Its Implications for Structure Growth}",
    eprint = "2304.05202",
    reportNumber = "FERMILAB-PUB-23-237-PPD, FERMILAB-PUB-23-237-PPD",
    doi = "10.3847/1538-4357/acfe06",
    journal = "\apj",
    volume = "962",
    number = "2",
    pages = "112",
    year = "2024"
}

@article{Abril-Cabezas:2025,
    author = "Abril-Cabezas, Irene and others",
    title = "{The Atacama Cosmology Telescope. CMB Lensing from Daytime Data: A First Demonstration}",
    eprint = "2511.10620",
    archivePrefix = "arXiv",
    primaryClass = "astro-ph.CO",
    month = "11",
    year = "2025"
}

@article{AliCPT:phi,
 author = "Liu, J. and Sun, Z. and Han, J. and Carron, J. and Delabrouille, J. and Li, S. and Liu, Y. and Jin, J. and Ghosh, S. and Yue, B. and Zhang, P. and Feng, C. and Huang, Z.-Q. and Liu, H. and Wu, Y.-W. and Zhang, L. and Zhang, Z.-R. and Zhao, W. and Hu, B. and Li, H. and Zhang, X.",
 title = "{Forecasts on CMB lensing observations with AliCPT-1}",
 eprint = "2204.08158",
 doi = "10.1007/s11433-022-1966-4",
 journal = "Sci. China Phys. Mech. Astron.",
 volume = "65",
 number = "10",
 pages = "109511",
 year = "2022"
}

@article{BKVIII,
 author = {{\textsc{Bicep2}/{\it Keck Array} Collaboration}},
 eprint = {1606.01968},
 journal= {\apj},
 pages  = {228},
 title  = {{\sc BICEP2}/{\it Keck Array} VIII: Measurement of gravitational lensing from large-scale B-mode polarization},
 volume = {833},
 year   = {2016},
}

@article{BICEPKeck:2022:los-dist,
 author = {{\textsc{Bicep2}/{\it Keck Array} Collaboration}},
 title = "{BICEP/Keck. XVII. Line-of-sight Distortion Analysis: Estimates of Gravitational Lensing, Anisotropic Cosmic Birefringence, Patchy Reionization, and Systematic Errors}",
 eprint = "2210.08038",
 doi = "10.3847/1538-4357/acc85c",
 journal = "\apj",
 volume = "949",
 number = "2",
 pages = "43",
 year = "2023"
}

@article{LiteBIRD,
 author = "{LiteBIRD Collaboration}",
 collaboration = "LiteBIRD",
 title = "{Probing Cosmic Inflation with the LiteBIRD Cosmic Microwave Background Polarization Survey}",
 eprint = "2202.02773",
 doi = "10.1093/ptep/ptac150",
 journal = "\ptep",
 volume = "2023",
 pages = "042F01",
 year = "2023"
}

@article{PR1:phi,
 author = {{\textit{Planck} Collaboration}},
 journal= {\aap},
 pages  = {A17},
 title  = {Planck 2013 results. XVII. Gravitational lensing by large-scale structure},
 volume = {571},
 year   = {2014},
 doi    = "10.1051/0004-6361/201321543",
 eprint = "1303.5077",
}

@article{PR2:phi,
 author = {{\textit{Planck} Collaboration}},
 eprint = {1502.01591},
 journal= {\aap},
 pages  = {A15},
 title  = {Planck 2015 results. XV. Gravitational lensing},
 volume = {594},
 year   = {2015},
}

@article{PR2:DE,
    author = "Ade, P. A. R. and others",
    collaboration = "Planck",
    title = "{Planck 2015 results. XIV. Dark energy and modified gravity}",
    eprint = "1502.01590",
    archivePrefix = "arXiv",
    primaryClass = "astro-ph.CO",
    doi = "10.1051/0004-6361/201525814",
    journal = "\aap",
    volume = "594",
    pages = "A14",
    year = "2016"
}

@article{PR3:phi,
 author = {{\textit{Planck} Collaboration} and {Aghanim}, N. and {Akrami}, Y. and
         {Ashdown}, M. and {Aumont}, J. and {Baccigalupi}, C. and
         {Ballardini}, M. and {Banday}, A.~J. and {Barreiro}, R.~B. and
         {Bartolo}, N. and {Basak}, S. and {Benabed}, K. and {Bernard}, J. -P. and
         {Bersanelli}, M. and {Bielewicz}, P. and {Bock}, J.~J. and
         {Bond}, J.~R. and {Borrill}, J. and {Bouchet}, F.~R. and
         {Boulanger}, F. and {Bucher}, M. and {Burigana}, C. and
         {Calabrese}, E. and {Cardoso}, J. -F. and {Carron}, J. and
         {Challinor}, A. and {Chiang}, H.~C. and {Colombo}, L.~P.~L. and
         {Combet}, C. and {Crill}, B.~P. and {Cuttaia}, F. and
         {de Bernardis}, P. and {de Zotti}, G. and {Delabrouille}, J. and
         {Di Valentino}, E. and {Diego}, J.~M. and {Dor{\'e}}, O. and
         {Douspis}, M. and {Ducout}, A. and {Dupac}, X. and {Efstathiou}, G. and
         {Elsner}, F. and {En{\ss}lin}, T.~A. and {Eriksen}, H.~K. and
         {Fantaye}, Y. and {Fernandez-Cobos}, R. and {Forastieri}, F. and
         {Frailis}, M. and {Fraisse}, A.~A. and {Franceschi}, E. and
         {Frolov}, A. and {Galeotta}, S. and {Galli}, S. and {Ganga}, K. and
         {G{\'e}nova-Santos}, R.~T. and {Gerbino}, M. and {Ghosh}, T. and
         {Gonz{\'a}lez-Nuevo}, J. and {G{\'o}rski}, K.~M. and {Gratton}, S. and
         {Gruppuso}, A. and {Gudmundsson}, J.~E. and {Hamann}, J. and {Hand
        ley}, W. and {Hansen}, F.~K. and {Herranz}, D. and {Hivon}, E. and
         {Huang}, Z. and {Jaffe}, A.~H. and {Jones}, W.~C. and {Karakci}, A. and
         {Keih{\"a}nen}, E. and {Keskitalo}, R. and {Kiiveri}, K. and {Kim}, J. and
         {Knox}, L. and {Krachmalnicoff}, N. and {Kunz}, M. and
         {Kurki-Suonio}, H. and {Lagache}, G. and {Lamarre}, J. -M. and
         {Lasenby}, A. and {Lattanzi}, M. and {Lawrence}, C.~R. and
         {Le Jeune}, M. and {Levrier}, F. and {Lewis}, A. and {Liguori}, M. and
         {Lilje}, P.~B. and {Lindholm}, V. and {L{\'o}pez-Caniego}, M. and
         {Lubin}, P.~M. and {Ma}, Y. -Z. and {Mac{\'\i}as-P{\'e}rez}, J.~F. and
         {Maggio}, G. and {Maino}, D. and {Mandolesi}, N. and {Mangilli}, A. and
         {Marcos-Caballero}, A. and {Maris}, M. and {Martin}, P.~G. and
         {Mart{\'\i}nez-Gonz{\'a}lez}, E. and {Matarrese}, S. and {Mauri}, N. and
         {McEwen}, J.~D. and {Melchiorri}, A. and {Mennella}, A. and
         {Migliaccio}, M. and {Miville-Desch{\^e}nes}, M. -A. and
         {Molinari}, D. and {Moneti}, A. and {Montier}, L. and {Morgante}, G. and
         {Moss}, A. and {Natoli}, P. and {Pagano}, L. and {Paoletti}, D. and
         {Partridge}, B. and {Patanchon}, G. and {Perrotta}, F. and
         {Pettorino}, V. and {Piacentini}, F. and {Polastri}, L. and
         {Polenta}, G. and {Puget}, J. -L. and {Rachen}, J.~P. and
         {Reinecke}, M. and {Remazeilles}, M. and {Renzi}, A. and {Rocha}, G. and
         {Rosset}, C. and {Roudier}, G. and {Rubi{\~n}o-Mart{\'\i}n}, J.~A. and
         {Ruiz-Granados}, B. and {Salvati}, L. and {Sandri}, M. and
         {Savelainen}, M. and {Scott}, D. and {Sirignano}, C. and {Sunyaev}, R. and
         {Suur-Uski}, A. -S. and {Tauber}, J.~A. and {Tavagnacco}, D. and
         {Tenti}, M. and {Toffolatti}, L. and {Tomasi}, M. and {Trombetti}, T. and
         {Valiviita}, J. and {Van Tent}, B. and {Vielva}, P. and {Villa}, F. and
         {Vittorio}, N. and {Wandelt}, B.~D. and {Wehus}, I.~K. and {White}, M. and
         {White}, S.~D.~M. and {Zacchei}, A. and {Zonca}, A.},
 eprint = {1807.06210},
 month = sep,
 volume = {641},
 eid = {A8},
 pages = {A8},
 doi = {10.1051/0004-6361/201833886},
 journal = {\aap},
 title  = {Planck 2018 results. VIII. Gravitational lensing},
 year   = {2020},
}

@article{PR3:main,
 author = {{\textit{Planck} Collaboration}},
 eprint = {1807.06209},
 title  = {Planck 2018 results. VI. Cosmological parameters},
 journal = {\aap},
 year   = {2018},
}

@article{Carron:2022:NPIPE-lensing,
 author = "Carron, J. and Mirmelstein, M. and Lewis, A.",
 title = "{CMB lensing from Planck PR4~maps}",
 eprint = "2206.07773",
 doi = "10.1088/1475-7516/2022/09/039",
 journal = "\jcap",
 volume = "09",
 pages = "039",
 year = "2022"
}

@article{PB:phi:2013,
 author = {{{\textsc POLARBEAR} Collaboration}},
 collaboration = {POLARBEAR},
 title = "{Measurement of the Cosmic Microwave Background Polarization Lensing Power Spectrum with the POLARBEAR experiment}",
 eprint = "1312.6646",
 doi = "10.1103/PhysRevLett.113.021301",
 journal = "\prl",
 volume = "113",
 pages = "021301",
 year = "2014"
}

@article{PB:phi:2019,
 author = {{{\textsc POLARBEAR} Collaboration}},
 collaboration = {POLARBEAR},
 title = "{Measurement of the Cosmic Microwave Background Polarization Lensing Power Spectrum from Two Years of POLARBEAR Data}",
 eprint = "1911.10980",
 doi = "10.3847/1538-4357/ab7e29",
 journal = "\apj",
 volume = "893",
 pages = "85",
 year = "2020"
}

@article{SimonsObservatory,
 author = "{Simons Observatory Collaboration}",
 title = "{The Simons Observatory: Science goals and forecasts}",
 eprint = "1808.07445",
 doi = "10.1088/1475-7516/2019/02/056",
 journal = "\jcap",
 volume = "02",
 pages = "056",
 year = "2019"
}

@article{SimonsObservatory:LAT,
 author = "{Simons Observatory Collaboration}",
 collaboration = "Simons Observatory",
 title = "{The Simons Observatory: science goals and forecasts for the enhanced Large Aperture Telescope}",
 eprint = "2503.00636",
 reportNumber = "FERMILAB-PUB-25-0188-PPD",
 doi = "10.1088/1475-7516/2025/08/034",
 journal = "\jcap",
 volume = "08",
 pages = "034",
 year = "2025"
}

@ARTICLE{SPT:phi:2012,
 author = {{van Engelen}, A. and {Keisler}, R. and {Zahn}, O. and {Aird}, K.~A. and {Benson}, B.~A. and {Bleem}, L.~E. and {Carlstrom}, J.~E. and {Chang}, C.~L. and {Cho}, H.~M. and {Crawford}, T.~M. and {Crites}, A.~T. and {de Haan}, T. and {Dobbs}, M.~A. and {Dudley}, J. and {George}, E.~M. and {Halverson}, N.~W. and {Holder}, G.~P. and {Holzapfel}, W.~L. and {Hoover}, S. and {Hou}, Z. and {Hrubes}, J.~D. and {Joy}, M. and {Knox}, L. and {Lee}, A.~T. and {Leitch}, E.~M. and {Lueker}, M. and {Luong-Van}, D. and {McMahon}, J.~J. and {Mehl}, J. and {Meyer}, S.~S. and {Millea}, M. and {Mohr}, J.~J. and {Montroy}, T.~E. and {Natoli}, T. and {Padin}, S. and {Plagge}, T. and {Pryke}, C. and {Reichardt}, C.~L. and {Ruhl}, J.~E. and {Sayre}, J.~T. and {Schaffer}, K.~K. and {Shaw}, L. and {Shirokoff}, E. and {Spieler}, H.~G. and {Staniszewski}, Z. and {Stark}, A.~A. and {Story}, K. and {Vanderlinde}, K. and {Vieira}, J.~D. and {Williamson}, R.},
 title = "{A Measurement of Gravitational Lensing of the Microwave Background Using South Pole Telescope Data}",
 journal = {\apj},
 year = 2012,
 volume = {756},
 number = {2},
 eid = {142},
 pages = {142},
 doi = {10.1088/0004-637X/756/2/142},
 eprint = {1202.0546},
}

@article{Story:2015,
 doi = {10.1088/0004-637X/810/1/50},
 url = {https://dx.doi.org/10.1088/0004-637X/810/1/50},
 year = {2015},
 month = {aug},
 volume = {810},
 number = {1},
 pages = {50},
 author = {K. T. Story and D. Hanson and P. A. R. Ade and K. A. Aird and J. E. Austermann and  J. A. Beall and A. N. Bender and B. A. Benson and L. E. Bleem and J. E. Carlstrom and C. L. Chang and H. C. Chiang and H-M. Cho and R. Citron and T. M. Crawford and A. T. Crites and T. de Haan and M. A. Dobbs and W. Everett and J. Gallicchio and J. Gao and E. M. George and A. Gilbert and N. W. Halverson and N. Harrington and J. W. Henning and G. C. Hilton and G. P. Holder and W. L. Holzapfel and S. Hoover and Z. Hou and J. D. Hrubes and N. Huang and J. Hubmayr and K. D. Irwin and R. Keisler and L. Knox and A. T. Lee and E. M. Leitch and D. Li and C. Liang and D. Luong-Van and J. J. McMahon and J. Mehl and S. S. Meyer and L. Mocanu and T. E. Montroy and T. Natoli and J. P. Nibarger and V. Novosad and S. Padin and C. Pryke and C. L. Reichardt and J. E. Ruhl and B. R. Saliwanchik and J. T. Sayre and K. K. Schaffer and G. Smecher and A. A. Stark and C. Tucker and K. Vanderlinde and J. D. Vieira and G. Wang and N. Whitehorn and V. Yefremenko and O. Zahn},
 title = {A MEASUREMENT OF THE COSMIC MICROWAVE BACKGROUND GRAVITATIONAL LENSING POTENTIAL FROM 100 SQUARE DEGREES OF SPTPOL DATA},
 journal = {\apj},
}

@article{SPT:phi:2019,
 author = {{Wu}, W.~L.~K. and {Mocanu}, L.~M. and {Ade}, P.~A.~R. and {Anderson}, A.~J. and {Austermann}, J.~E. and {Avva}, J.~S. and {Beall}, J.~A. and {Bender}, A.~N. and {Benson}, B.~A. and {Bianchini}, F. and {Bleem}, L.~E. and {Carlstrom}, J.~E. and {Chang}, C.~L. and {Chiang}, H.~C. and {Citron}, R. and {Corbett Moran}, C. and {Crawford}, T.~M. and {Crites}, A.~T. and {de Haan}, T. and {Dobbs}, M.~A. and {Everett}, W. and {Gallicchio}, J. and {George}, E.~M. and {Gilbert}, A. and {Gupta}, N. and {Halverson}, N.~W. and {Harrington}, N. and {Henning}, J.~W. and {Hilton}, G.~C. and {Holder}, G.~P. and {Holzapfel}, W.~L. and {Hou}, Z. and {Hrubes}, J.~D. and {Huang}, N. and {Hubmayr}, J. and {Irwin}, K.~D. and {Knox}, L. and {Lee}, A.~T. and {Li}, D. and {Lowitz}, A. and {Manzotti}, A. and {McMahon}, J.~J. and {Meyer}, S.~S. and {Millea}, M. and {Montgomery}, J. and {Nadolski}, A. and {Natoli}, T. and {Nibarger}, J.~P. and {Noble}, G.~I. and {Novosad}, V. and {Omori}, Y. and {Padin}, S. and {Patil}, S. and {Pryke}, C. and {Reichardt}, C.~L. and {Ruhl}, J.~E. and {Saliwanchik}, B.~R. and {Sayre}, J.~T. and {Schaffer}, K.~K. and {Sievers}, C. and {Simard}, G. and {Smecher}, G. and {Stark}, A.~A. and {Story}, K.~T. and {Tucker}, C. and {Vanderlinde}, K. and {Veach}, T. and {Vieira}, J.~D. and {Wang}, G. and {Whitehorn}, N. and {Yefremenko}, V.},
 title = "{A Measurement of the Cosmic Microwave Background Lensing Potential and Power Spectrum from 500 deg$^{2}$ of SPTpol Temperature and Polarization Data}",
 journal = {\apj},
 year = 2019,
 month = oct,
 volume = {884},
 number = {1},
 pages = {70},
 doi = {10.3847/1538-4357/ab4186},
 eprint = {1905.05777},
}

@article{Millea:2020,
 author = "{Millea}, M. and {Daley}, C.~M. and {Chou}, T.-L. and {Anderes}, E. and {Ade}, P.~A.~R. and {Anderson}, A.~J. and {Austermann}, J.~E. and {Avva}, J.~S. and {Beall}, J.~A. and {Bender}, A.~N. and {Benson}, B.~A. and {Bianchini}, F. and {Bleem}, L.~E. and {Carlstrom}, J.~E. and {Chang}, C.~L. and {Chaubal}, P. and {Chiang}, H.~C. and {Citron}, R. and {Moran}, C. Corbett and {Crawford}, T.~M. and {Crites}, A.~T. and {de Haan}, T. and {Dobbs}, M.~A. and {Everett}, W. and {Gallicchio}, J. and {George}, E.~M. and {Goeckner-Wald}, N. and {Guns}, S. and {Gupta}, N. and {Halverson}, N.~W. and {Henning}, J.~W. and {Hilton}, G.~C. and {Holder}, G.~P. and {Holzapfel}, W.~L. and {Hrubes}, J.~D. and {Huang}, N. and {Hubmayr}, J. and {Irwin}, K.~D. and {Knox}, L. and {Lee}, A.~T. and {Li}, D. and {Lowitz}, A. and {McMahon}, J.~J. and {Meyer}, S.~S. and {Mocanu}, L.~M. and {Montgomery}, J. and {Natoli}, T. and {Nibarger}, J.~P. and {Noble}, G. and {Novosad}, V. and {Omori}, Y. and {Padin}, S. and {Patil}, S. and {Pryke}, C. and {Reichardt}, C.~L. and {Ruhl}, J.~E. and {Saliwanchik}, B.~R. and {Schaffer}, K.~K. and {Sievers}, C. and {Smecher}, G. and {Stark}, A.~A. and {Thorne}, B. and {Tucker}, C. and {Veach}, T. and {Vieira}, J.~D. and {Wang}, G. and {Whitehorn}, N. and {Wu}, W.~L.~K. and {Yefremenko}, V.",
 title = "{Optimal Cosmic Microwave Background Lensing Reconstruction and Parameter Estimation with SPTpol Data}",
 eprint = "2012.01709",
 reportNumber = "FERMILAB-PUB-20-676-AE",
 doi = "10.3847/1538-4357/ac02bb",
 month = dec,
 journal = "\apj",
 volume = "922",
 number = "2",
 pages = "259",
 year = "2021"
}

@article{Pan:SPT:2023,
 author = "{Pan}, Z. and {Bianchini}, F. and {Wu}, W.~L.~K. and {Ade}, P.~A.~R. and {Ahmed}, Z. and {Anderes}, E. and {Anderson}, A.~J. and {Ansarinejad}, B. and {Archipley}, M. and {Aylor}, K. and {Balkenhol}, L. and {Barry}, P.~S. and {Basu Thakur}, R. and {Benabed}, K. and {Bender}, A.~N. and {Benson}, B.~A. and {Bleem}, L.~E. and {Bouchet}, F.~R. and {Bryant}, L. and {Byrum}, K. and {Camphuis}, E. and {Carlstrom}, J.~E. and {Carter}, F.~W. and {Cecil}, T.~W. and {Chang}, C.~L. and {Chaubal}, P. and {Chen}, G. and {Chichura}, P.~M. and {Cho}, H.-M. and {Chou}, T.-L. and {Cliche}, J.-F. and {Coerver}, A. and {Crawford}, T.~M. and {Cukierman}, A. and {Daley}, C. and {de Haan}, T. and {Denison}, E.~V. and {Dibert}, K.~R. and {Ding}, J. and {Dobbs}, M.~A. and {Doussot}, A. and {Dutcher}, D. and {Everett}, W. and {Feng}, C. and {Ferguson}, K.~R. and {Fichman}, K. and {Foster}, A. and {Fu}, J. and {Galli}, S. and {Gambrel}, A.~E. and {Gardner}, R.~W. and {Ge}, F. and {Goeckner-Wald}, N. and {Gualtieri}, R. and {Guidi}, F. and {Guns}, S. and {Gupta}, N. and {Halverson}, N.~W. and {Harke-Hosemann}, A.~H. and {Harrington}, N.~L. and {Henning}, J.~W. and {Hilton}, G.~C. and {Hivon}, E. and {Holder}, G.~P. and {Holzapfel}, W.~L. and {Hood}, J.~C. and {Howe}, D. and {Huang}, N. and {Irwin}, K.~D. and {Jeong}, O. and {Jonas}, M. and {Jones}, A. and {K{\'e}ruzor{\'e}}, F. and {Khaire}, T.~S. and {Knox}, L. and {Kofman}, A.~M. and {Korman}, M. and {Kubik}, D.~L. and {Kuhlmann}, S. and {Kuo}, C.-L. and {Lee}, A.~T. and {Leitch}, E.~M. and {Levy}, K. and {Lowitz}, A.~E. and {Lu}, C. and {Maniyar}, A. and {Menanteau}, F. and {Meyer}, S.~S. and {Michalik}, D. and {Millea}, M. and {Montgomery}, J. and {Nadolski}, A. and {Nakato}, Y. and {Natoli}, T. and {Nguyen}, H. and {Noble}, G.~I. and {Novosad}, V. and {Omori}, Y. and {Padin}, S. and {Paschos}, P. and {Pearson}, J. and {Posada}, C.~M. and {Prabhu}, K. and {Quan}, W. and {Raghunathan}, S. and {Rahimi}, M. and {Rahlin}, A. and {Reichardt}, C.~L. and {Riebel}, D. and {Riedel}, B. and {Ruhl}, J.~E. and {Sayre}, J.~T. and {Schiappucci}, E. and {Shirokoff}, E. and {Smecher}, G. and {Sobrin}, J.~A. and {Stark}, A.~A. and {Stephen}, J. and {Story}, K.~T. and {Suzuki}, A. and {Takakura}, S. and {Tandoi}, C. and {Thompson}, K.~L. and {Thorne}, B. and {Trendafilova}, C. and {Tucker}, C. and {Umilta}, C. and {Vale}, L.~R. and {Vanderlinde}, K. and {Vieira}, J.~D. and {Wang}, G. and {Whitehorn}, N. and {Yefremenko}, V. and {Yoon}, K.~W. and {Young}, M.~R. and {Zebrowski}, J.~A.",
 eprint = "2308.11608",
 primaryClass = "astro-ph.CO",
 title = "{Measurement of gravitational lensing of the cosmic microwave background using SPT-3G 2018 data}",
 journal = {\prd},
 month = dec,
 volume = {108},
 number = {12},
 eid = {122005},
 pages = {122005},
 doi = {10.1103/PhysRevD.108.122005},
 year = "2023"
}

@article{Ge:SPT-3G:2024,
    author = "Ge, F. and others",
    collaboration = "SPT-3G",
    title = "{Cosmology from CMB lensing and delensed EE power spectra using 2019{\textendash}2020 SPT-3G polarization data}",
    eprint = "2411.06000",
    archivePrefix = "arXiv",
    primaryClass = "astro-ph.CO",
    reportNumber = "FERMILAB-PUB-24-0840-PPD",
    doi = "10.1103/PhysRevD.111.083534",
    journal = "Phys. Rev. D",
    volume = "111",
    number = "8",
    pages = "083534",
    year = "2025"
}

\end{document}